\documentclass[aps,nobibnotes,twocolumn,superscriptaddress,showkeys,thelongbibliography]{revtex4-2}   	
\usepackage{graphicx}				
											
\usepackage{amssymb}
\usepackage{amsmath}
\usepackage{hyperref}
\usepackage{multirow}
\usepackage{placeins}
\usepackage{etoolbox}

\newtoggle{arxiv}
\toggletrue{arxiv} 

\begin{document}
\newcommand{\eelsix}{Video A}
\newcommand{\eelfour}{Video B}



\title{
Metachronal waves in concentrations of swimming {\it Turbatrix aceti} nematodes
and an oscillator chain model for their coordinated motions}

\author{A. C. Quillen}
\email{alice.quillen@rochester.edu}
\affiliation{Department of Physics and Astronomy, University of Rochester, Rochester, NY 14627, USA}
\author{A. Peshkov}
\email{apeshkov@ur.rochester.edu}
\affiliation{Department of Physics and Astronomy, University of Rochester, Rochester, NY 14627, USA}
\author{Esteban Wright}
\email{ewrig15@ur.rochester.edu }
\affiliation{Department of Physics and Astronomy, University of Rochester, Rochester, NY 14627, USA}
\author{Sonia McGaffigan}
\email{smcgaffi@u.rochester.edu }
\affiliation{Department of Physics and Astronomy, University of Rochester, Rochester, NY 14627, USA}

\begin{abstract}
At high concentration, free swimming nematodes known as vinegar eels ({\it Turbatrix aceti}), collectively exhibit metachronal waves near a boundary.  We find that the frequency of the collective traveling wave is lower than that of the freely swimming organisms. We explore models based on a chain of oscillators with nearest neighbor interactions that inhibit oscillator phase velocity. The phase of each oscillator represents the phase of the motion of the eel's head back and forth about its mean position.  A strongly interacting directed chain model mimicking steric repulsion between organisms robustly gives traveling wave states and can approximately match the observed wavelength and oscillation frequency of the observed traveling wave.  We predict body shapes assuming that waves propagate down the eel body at a constant speed.  The phase oscillator model that impedes eel head overlaps also reduces close interactions throughout the eel bodies.   
 \end{abstract}



\maketitle

\section{Introduction}
\label{sec:intro}

Concentrations of biological organisms can be considered active materials as they are comprised
of self-driven units and energy is
continuously expended through locomotion \citep{Marchetti_2013}.
Collective behavior of groups of organisms include flocking or swimming in schools  \citep{Partridge_1982,Calovi_2014}  and synchronization \citep{Buck_1966,Strogatz_2012}. 
Synchronization processes in nature include glowing rhythms of colonies of fireflies \citep{Buck_1966}, 
crowd synchrony of pedestrians walking on a bridge \citep{Strogatz_2005} 
and flagella beating in phase with one another  \citep{Taylor_1951}.

The head or tail of an individual snake, flagellum, cilium or nematode moves back and forth
with respect to a mean position.  This periodic motion 
can be described with a phase of oscillation (e.g., \citep{Uchida_2011}). 
In concentrations of mobile oscillators, both synchronization and swarming can occur together, 
and such systems can display a rich diversity of collective states 
(e.g., the {\it swarmalators}  studied by  \citet{OKeeffe_2017})
including collectively organized and coordinated motions known as 
traveling or metachronal waves. 
A metachronal rhythm or metachronal wave refers to a locally synchronized motion of individuals with a delay between them, in contrast to globally synchronized patterns of oscillation. 

Metachronal waves require coordinated motions between neighboring structures or organisms
\citep{Winfree_2002,Lenz_2006}.
Swimming spermatozoa synchronize the beating of their cilia, and flagellates 
 can synchronize the motions of their flagella when they are in close proximity \citep{Taylor_1951,Tamm_1975,Niedermayer_2008,Brumley_2012,Elgeti_2013,Elgeti_2015}.   
When a constant phase difference or time delay is maintained between neighboring  
oscillating structures,  the collective motion has the appearance of a traveling wave.

One approach to modeling metachronal wave formation 
in cilia or flagella is to model them as an array of flexible filaments that oscillate or beat when alone.
Self-organized metachronal waves then arise due to hydrodynamic 
\citep{Gueron_1998,Vilfan_2006,Lenz_2006,Niedermayer_2008,Lindemann_2010,Uchida_2011,Uchida_2012,Brumley_2012}
or steric \citep{Chelakkot_2021} interactions between neighboring filaments.
Even though a filament can bend and flex, its behavior can approximately
be described with an angle or phase which specifies the position of its moving tip 
(e.g., \citep{Lenz_2006,Brumley_2012,Niedermayer_2008}).  Although each filament moves
in three dimensions, simplified models consisting of discrete linear chains of interacting oscillators 
can describe the collective behavior \citep{Lenz_2006,Brumley_2012,Niedermayer_2008}.

Phase oscillator chain models, known as local Kuramoto models, exhibit both long lived synchronous and traveling wave states \citep{Ermentrout_1986,Ermentrout_1990,Ren_2000,Muruganandam_2008,Tilles_2011,Denes_2019}.   However, in many of these models, a system with randomly chosen initial phases is more likely to evolve into a synchronous state than a traveling wave state \citep{Tilles_2011,Denes_2019}.
Simple criteria are not available for predicting whether an interacting phase oscillator model is likely to give traveling wave states if initialized with random phases.  However, physically motivated interacting phase oscillator models for metachronal waves in cilia and flagella have succeeded in robustly giving traveling wave states  \citep{Brumley_2012,Niedermayer_2008}. 

In this study we report on collective behavior in a system of undulating free-swimming organisms, vinegar eels, 
species {\it Turbatrix aceti} ({\it T. aceti}),  which are a type of free-living nematode. 
They are found living in beer mats, slime from tree wounds and cultures of edible vinegars.
Because they are hardy, they are used in aquaculture by fish keepers and aquarists as food for newly hatched fish or crustaceans.  Vinegar eels are tolerant of variation in acidity and they subsist on yeast.
The metachronal waves in {\it T. aceti}, reported by \citet{Peshkov_2019,Peshkov_2021}, are similar
to those seen in cilia.  However,  unlike cilia which are affixed to a cell membrane, the vinegar eels
are freely swimming organisms.   At about 1\ mm in length, the vinegar eels are visible by naked eye
and are much larger than cilia (typically a few $\mu$m in length) 
or flagella on colonies of microorganisms that display metachronal waves 
(e.g., with flagella length $\sim 10 \mu$m; \citep{Brumley_2012}).  Concentrated suspensions of
vinegar eels are a novel biological 
system in which we can study ensemble coordination and synchronization.
Henceforth we refer to the vinegar eels colloquially as `eels' even though they are nematodes.

Ensembles of active particles can exhibit a phase transition from gaseous to collective behavior at higher number density due to particle interactions (e.g., for unipolar self-propelled particles \citep{Vicsek_1995}).     
Metachronal waves are only present in high concentrations of vinegar eels \citep{Peshkov_2021} 
so interactions between them are necessary for the coordinated wave motion.   
Collective coordinated motion is likely to be mediated by the interactions between
the organisms.   In our study we compare the motion of the vinegar eels participating in metachronal waves
to those that are freely swimming to probe the nature of these interactions. 

While the well studied nematode {\it Caenorhabditis elegans} ({\it C. elegans}) naturally grows in soil, 
{\it C. elegans} is also an undulatory swimmer in water (e.g., \citep{Yuan_2015}).  {\it C. elegans} nematodes congregate near surfaces and boundaries (they exhibit bordertaxis) \citep{Yuan_2015}.    
In close proximity, a pair of swimming {\it C. elegans} nematodes will synchronize their gait \citep{Yuan_2014}. 
Collective behavior of {\it C. elegans} includes the formation of a network on a surface 
\citep{Sugi_2019} and synchronization of clusters of tens of nematodes \citep{Yuan_2014}.
We have observed 
 similarities between the reported behavior of {\it C. elegans} and our vinegar eel nematodes. 
These similarities include undulatory swimming, bordertaxis, and synchronization in the gait of clusters of organisms.
We have not found descriptions of metachronal waves in concentrations of {\it C. elegans} 
 or other nematodes in the literature 
nor have we seen metachronal waves in concentrations of {\it C. elegans} in our lab
\citep{Peshkov_2021}.


We briefly describe our experimental methods
in \ref{sec:methods}.   Measurements of individual vinegar eels
 at low concentration are discussed in section \ref{sec:low}. 
We describe the behavior of high concentrations of vinegar eels in section \ref{sec:high}.
Models of metachronal waves in cilia and flagella have described 
these systems as a chain of interacting phase oscillators, where each phase describes the motion of a cilium or
flagellum tip \citep{Niedermayer_2008,Brumley_2012}.  In section \ref{sec:osc} we adopt a similar
approach and model our ensemble of vinegar eels with a chain of interacting oscillators, but each  
 phase describes the motion of an eel's head.
A summary and discussion follows in section \ref{sec:sum}.

\section{Experimental Methods}
\label{sec:methods}

We obtained our {\it T. aceti} nematode and yeast culture from an aquarium supply store, and we grow it at room temperature in a 1:1 mixture of water and food grade apple cider vinegar.  A few slices of apple were added to the mixture as a food source for the yeast.   After a few ml of the purchased culture is added to the vinegar and apple mixture, it takes a few weeks before large numbers of vinegar eels are visible by eye in the mixture.  The vinegar eels congregate at the surface and crawl up the container walls.

To study the motion of the vinegar eels, we used a 
Krontech Chronos 1.4 high speed video camera at 1057 frames per second (fps)
giving image frames with 1024 $\times$ 1280 pixels.
To connect the video camera to a conventional stereo compound microscope under bright field illumination, 
we used a 0.5X reduction lens adapter that matches 
the C-mount of our camera. The other end of the adapter fits in
the 23.2\ mm diameter eyepiece holder of our microscope. 
Videos were taken using the X4 or X10 microscope objectives.

At each magnification, we made short videos of a calibration slide with a small ruler
on it.  Frames from these videos were used to measure the pixel scale,  
giving 315 mm/pixel and 838 mm/pixel at X4 and X10 magnification, respectively.
The field of view is 1.22 mm $ \times$ 1.53 mm at X4  magnification  and  3.25 mm $\times$ 4.06 mm
at  X10 magnification.

We present two videos,  both taken on Feb 26, 2020. 
The first video \cite{videoA}, denoted \eelsix,  filmed at X10 magnification, is of the vinegar eels at low concentration.   
The second video \cite{videoB}, denoted \eelfour,  is at higher concentration and was filmed at X4 magnification.
To achieve high vinegar eel concentration, 
we placed about 10 ml of the vinegar eel culture in a test tube
and then used a centrifuge (a few minutes at a few thousand rpm 
or about 1000 $g$) to concentrate the eels at the bottom. 
A pipette was then used to extract fluid from the bottom of the tube.  

Each video views a drop  of 
about $100 \mu$l of dilute vinegar containing vinegar eels that was deposited on a dry 
glass slide.  The drop was not covered with a coverslip, so its surface is curved 
due to surface tension.  The slides wet so the drop is not spherical.  The outer edge of the drop
where it touches the slide remains fixed due to surface tension.  
In both videos, the drop was about a cm in diameter.
In \eelfour, we touched the edge of the drop with a metal pin a few times to pull and extend the drop radially outward.
This increased the drop surface area on the slide and decreased its depth.
This system is nearly two dimensional as the vinegar eels rarely swim above or below one another.  
Additional experiments of drops containing {\it T. aceti} are discussed by \citet{Peshkov_2021}.

\begin{figure} 
\begin{tabular}{c} 
\includegraphics[width=3.5in, trim=10 0 10 10, clip]{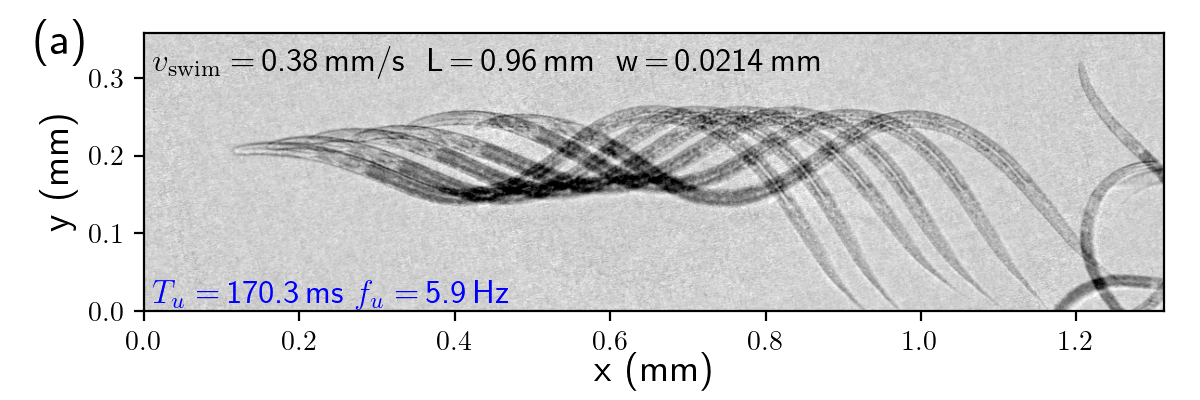}\\
\includegraphics[width=3.5in, trim=10 0 10 30, clip]{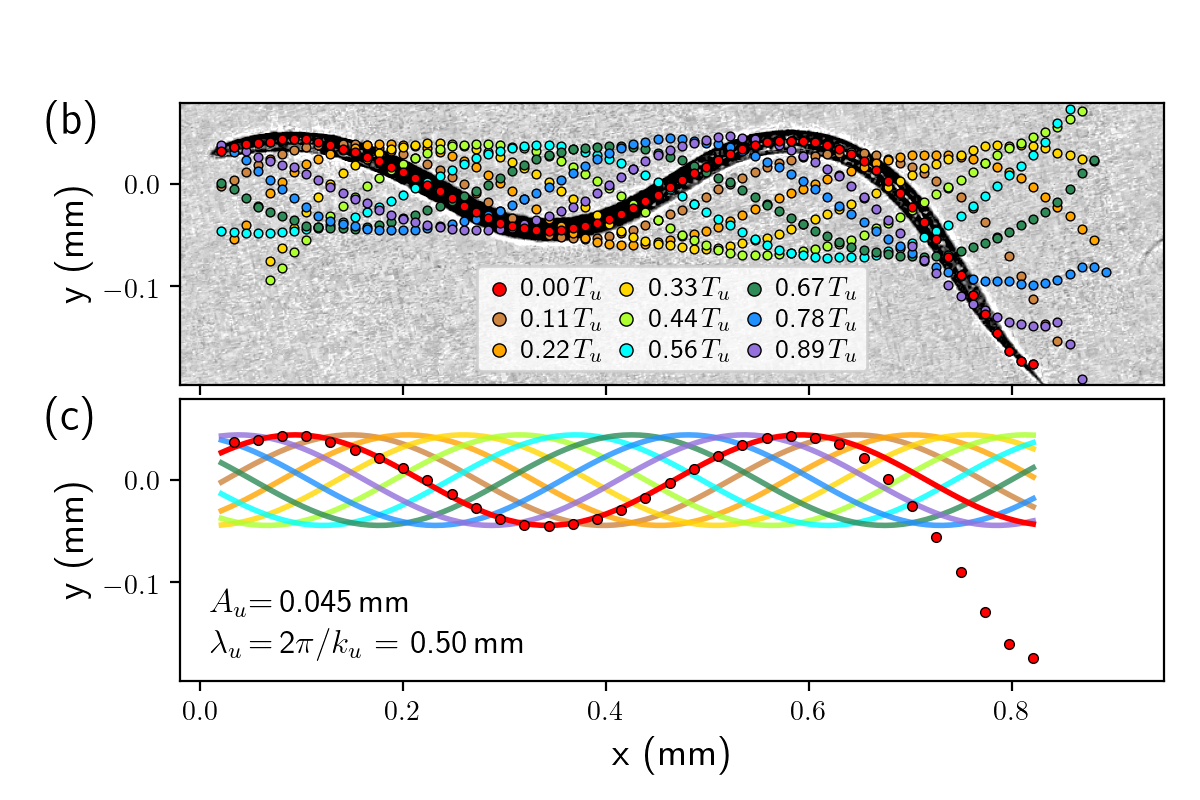}
\end{tabular}
\caption{Characteristics of an adult mm long freely-swimming vinegar eel.
(a) The gray-scale image shows a sum of 5 frames from high speed \eelsix \ showing the same freely swimming
 eel.  The 5 frames are equally spaced in time with  interval $T_u=170.3$ ms 
that is approximately one oscillation or undulation period. 
The oscillation frequency is written on the lower left in Hz; $f_u  = 1/T_u$.
The swim speed, $v_{\rm swim}$, eel length, $L$  and diameter $w$ are written on the top of the frame.
The images have been rotated so that the organism is swimming in the horizontal direction and to the left.
(b) Body positions are shown with
colored dots at 9 equally spaced times during a single oscillation period. 
The images used to measure the body position have been shifted to take into account
the mean swim speed. 
The body positions are plotted on top of the first video frame in the sequence.
(c) Using the first time shown in (b), 
the $y$ position of the center of the eel body as a function of $x$ is plotted with red dots
enclosed in black circles.
The red line shows the sine function $y = A_u \cos( k_ux - \phi_0)$ fit to these points.  
The wavelength and amplitude
of this function are shown on the lower left.  
The colored lines show $y = A_u \cos (k_u x - \phi_0 - 2\pi j/9)$ for integers $j \in 1 ... 8 $ corresponding
to the phases of oscillation shown in (b).
The eel body is approximately sinusoidal in shape over much of its body and during most of its gait.
\label{fig:eel6_large}}
\end{figure}

\begin{table*}
\caption{Properties of a freely swimming vinegar eel \label{tab:free}}
\begin{tabular}{llllllll}
\hline
Quantity  & Symbol & units    &    Value    \\
\hline
Length                &   $L$          &   mm     &  $0.96 \pm 0.03$   \\
Diameter             & $w$           & mm     & $0.021 \pm 0.001$  \\
Length/diameter  & $L/w$       &   -        & 45  \\
Wavelength         & $\lambda_u$ &  mm   &  $0.50 \pm 0.02$   \\
Amplitude            &   $A_u$         & mm  &  $0.045 \pm 0.005$        \\
Swim speed        & $v_{\rm swim}$  & mm/s & $0.38\pm 0.03$   \\ 
Amplitude/phys. length    &   $A_u/h_x$   & -  &  0.055       \\
Amplitude times wave-vector   &  $A_u k_u$    & - & 0.56 \\
Oscillation period           &  $T_u$   & ms & $170 \pm 6$     \\
Oscillation frequency     & $f_u=1/T_u$  &  Hz & $5.9\pm 0.2$  \\
Undulation wave speed along body & $v_u=\lambda_u/T_u$ & mm/s & 3.0  \\
\hline
\end{tabular}\\
The  length $h_x$ is the linear distance between head and tail measured along the direction of motion.
The length $L$ is that of the eel, integrated along its body or measured if it were extended to its maximum length.
Because the eel is not straight while it is swimming $h_x<L$.
The wave speed along the body is that of undulation.  Uncertainties describe 
the range of values that would be consistent with the motion during a 1 s long
segment of video.   The vinegar eel is shown in Figure \ref{fig:eel6_large}.
\end{table*}

\section{Observations of lone eels at low concentration}
\label{sec:low}

In \eelsix, the vinegar eels are at low concentration and we can find intervals when
an individual eel is not strongly influenced by nearby eels or borders.  
We focus on an adult  $\sim 1$ mm long vinegar eel, shown in Figure \ref{fig:eel6_large}, 
 because it can be directly compared to
prior work studying 1\ mm long {\it C. elegans} kinematics (e.g., \citep{Sznitman_2010,Wen_2012,Yuan_2015})
and because eels of this length actively participate in the metachronal wave.

A median image was subtracted from all frames in \eelsix \ to remove smooth variations in lighting. 
After subtracting the median image, 
we rotated the video frames so that the lone vinegar eel swims to the left.
To find the eel's oscillation or gait period we summed 5 equally spaced (in time) video frames. 
We adjusted the time interval between the frames until the eel body shape was similar in each of the 5 frames,
indicating that they are at about the same phase of undulation.
This time interval gives us an estimate for the eel undulation period $T_u$.
The sum of 5 images is shown in Figure \ref{fig:eel6_large}a with the eel head on the left.
    
We estimated the eel's mean swim speed, $v_{\rm swim}$ by 
shifting the images so that the eel bodies in the 5 video frames 
appear to be at the same position.  The required shift to align
the eels after one oscillation period divided by the oscillation period $T_u$ gives 
the mean swim speed, $v_{\rm swim}$.

We used the mean swim speed to shift the video images so that positions are
viewed in the reference frame moving with this average speed. 
At 9 different phases of oscillation during a single oscillation period, we measured eel body centerlines
by fitting Gaussian functions to equally spaced vertical slices in the image. 
The mean of the Gaussian gives the eel's centerline $y$ value as a function of horizontal distance $x$.   
The body centerlines at these 9 different phases of oscillation
 are shown with different colored dots in Figure \ref{fig:eel6_large}b.
 The body centerlines are plotted on top of the first 
 video frame in the sequence which is shown with the underlying  grayscale image.  
In this figure, the origin is near the head's mean position. The positive $x$ axis opposite to the swim
direction and the $y$ axis is perpendicular to it.    
 
By integrating distances between the 
 points along the eel's centerline, we computed the length $L$ of the eel. 
We measured the eel's body diameter $w$ by measuring its apparent width 
across its middle.    In Figure \ref{fig:eel6_large}b the horizontal extent of the eel 
$h_x$ along the $x$ axis is smaller than the eel length because the eel body is not straight.

To estimate a beat amplitude $A_u$ and a wave vector $k_u$,
we fit a sine wave to the body centerline at one phase of oscillation
\begin{equation}
 y(x) =  A_u \cos (k_ux  - \phi_0) .\end{equation}
Figure \ref{fig:eel6_large}c shows the fit sine function with a red line. 
The sine describes the $y$ coordinate of the eel's centerline as a function of $x$
and $\phi_0$ is a phase.
The wavelength of the body shape $\lambda_u = 2 \pi/k_u$.
The amplitude $A_u$ describes the size of deviations from the mean of the centerline.
The speed that waves travel down the body $v_u$ is estimated from $v_u = \lambda_u/T_u $.

Measurements of the freely swimming vinegar eel are summarized in Table \ref{tab:free}.
Uncertainties listed in this table give the range of values that are consistent 
with the eel's motion during a 1 s long segment  of video.

The centerline positions in Figure \ref{fig:eel6_large}b  
show that larger amplitude motions, or larger deviations from a pure sine
shape occur at the head and tail of the vinegar eel.   
 Over much of the body the eel's shape is well described
with a sine function and the eel's body is nearly sinusoidal in shape 
during most of its oscillation.   The spacing and offsets 
between centerline curves at different phases of oscillation in Figure \ref{fig:eel6_large}b and c
imply the advance of the sine shape occurs at a nearly constant wave speed.



\begin{figure}  
\includegraphics[width=3.45in, trim=60 95 50 50, clip]{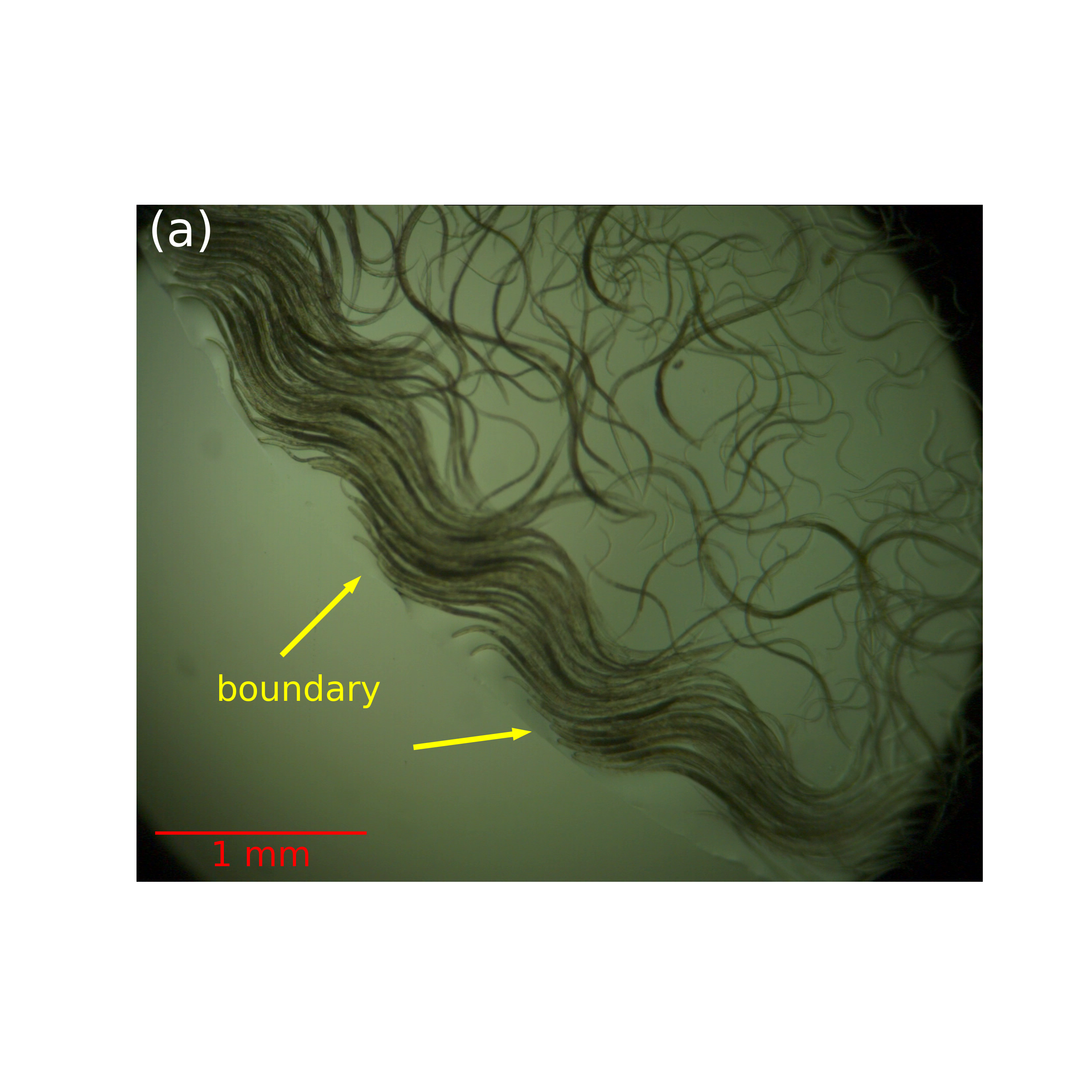}
\includegraphics[width=3.45in, trim=0 0 0 0, clip]{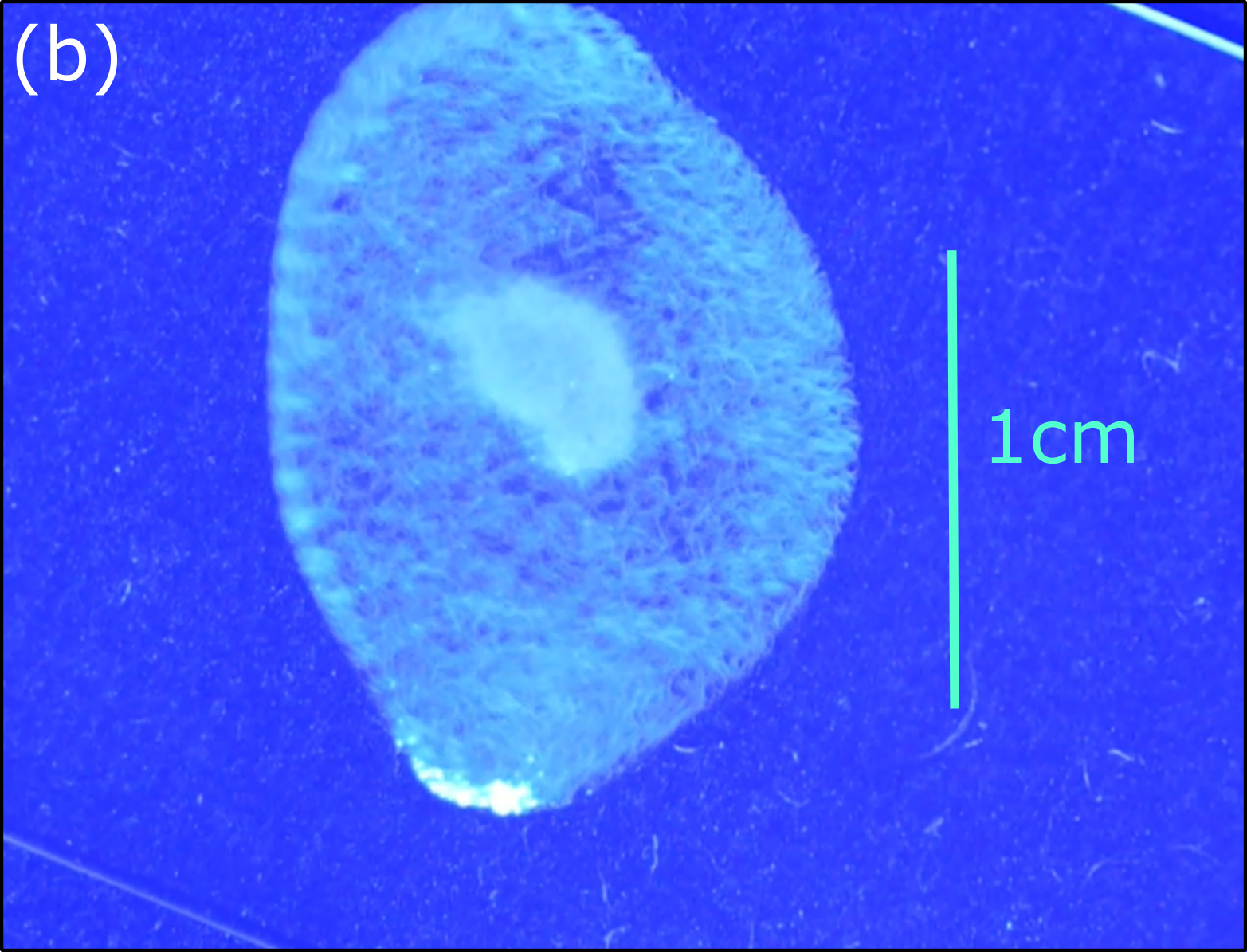}
\includegraphics[width=3.45in, trim=0 0 0 0, clip]{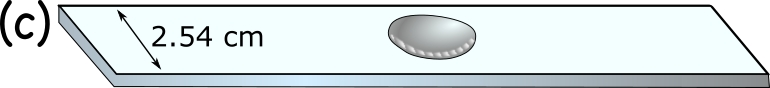}
\caption{(a) A raw video frame from \eelfour.    This video  
is of a dilute vinegar drop containing a high concentration of vinegar eels
seen through a conventional microscope at X4 magnification.
The edge of the drop on the slide is marked with yellow arrows.
The concentration of eels is higher near the edge of the drop.
There are eels of different lengths and ages in the solution, however  the smaller eels
are less likely to participate in the metachronal wave. 
(b) A photograph taken from above of a drop on a slide containing a high concentration of vinegar eels.
Detritus in the culture has been pushed to the center of the drop.
The feathery white ridges on the edge of the drop are the metachronal wave.
(c) An illustration of the drop of concentrated vinegar eel solution on a slide.
The white feathery features represent the traveling wave
in the vinegar eels near the edge of the drop.
\label{fig:wide}}
\end{figure}

Our vinegar eels culture contains nematodes of different sizes,  ranging from about 0.3 to 2 mm in length
(see Figure \ref{fig:wide}a).  We measured the frequency of oscillation for different length
eels and found that this frequency is not strongly dependent on eel length. We have noted 
that the ratio of length to wavelength $L/\lambda_u$ is larger for the larger and longer eels than the smaller ones.  
In the longer eels about 1.5 wavelengths are present whereas only 1 wavelength is present
on the shorter ones. 

The key findings of this section are the measurement of
the frequency of undulation for freely swimming vinegar eels ($f_u \sim 6 $ Hz) and that the 
shape and motion of much of the vinegar eel's body can be described with a sine function. 

\subsection{Comparison between {\it C. elegans} and {\it  T. aceti}}
\label{sec:comp}

Since the {\it C. elegans} nematode is well studied, we compare its kinematics to that of
the vinegar eel nematode,  {\it T. aceti}.
The frequency of undulation we measured in the vinegar eels $\sim 6$ Hz is  
faster than the $\sim 2$ Hz measured  in similar length (1 mm long) 
{\it C. elegans} \citep{Sznitman_2010,Yuan_2015}.   The length to diameter ratio for our 1 mm eel is about 
$L/w \sim$ 45 whereas {\it C. elegans} is not as slender with $L/w \sim 12$  \citep{Sznitman_2010}. 
More than 1 wavelength fits within the eel body in {\it T. aceti}, particularly in the longer eels.
In contrast about a single wavelength fits on the {\it C. elegans} body while it is swimming  \citep{Sznitman_2010}.
The speed that waves travel down the body, $v_u \sim 3 $ mm/s for the eel,  is somewhat higher than 
than of {\it C. elegans} (2.1 mm/s, \citep{Sznitman_2010}).  The swim speeds 
are similar;   0.4 mm/s for the 1 mm long vinegar eel  and 0.36 mm/s in {\it  C. elegans}.

In the vinegar eels, the amplitude of motion is larger at the head and tail, than in the middle
and is largest at the tail.  This behavior is similar
to swimming {\it C. elegans} \citep{Yuan_2015}  (see their Figure 1a) 
though  \citet{Sznitman_2010} measured the largest
body curvature variations near the head.

For vinegar eels at low concentration,
we did not find a significant difference between the undulation frequency 
of eels that are swimming near or along the edge of the drop
and of those that are swimming in the center of a drop.  In this respect
our vinegar eels are similar to {\it C. elegans}.  
For  {\it C. elegans} exhibiting bordertaxis and swimming
near a surface, the frequency 
of oscillation is  similar to that of the freely swimming organism \citep{Yuan_2015}.

\begin{table*}
\caption{Metachronal wave measurements \label{tab:meta}}
\begin{tabular}{llllllll}
\hline
Quantity  & Symbol &  Value        \\
\hline
Metachronal wave velocity          &   $v_{\rm MW}$        &      3.7 $\pm$ 0.2 mm/s     \\
Metachronal wave frequency       &   $f_{\rm MW}$        &       4.0 $\pm$ 0.2   Hz    \\
Wavelength of metachronal wave  &   $\lambda_{\rm MW}$     &  0.89 $\pm$ 0.03    mm  \\
Number of eels per wavelength     & $N_{\rm MW}$             & 13-16 \\
Ratio of frequencies                       & $f_{\rm MW}/f_u$   & $\sim  0.68$  \\
Amplitude of motion                       & $A_{\rm MW} $      & $ \sim 0.07 $ mm \\
\hline
\end{tabular}  
\end{table*}                     

\begin{figure} 
\centering\includegraphics[width=3.3in, trim=60 60 10 0, clip]{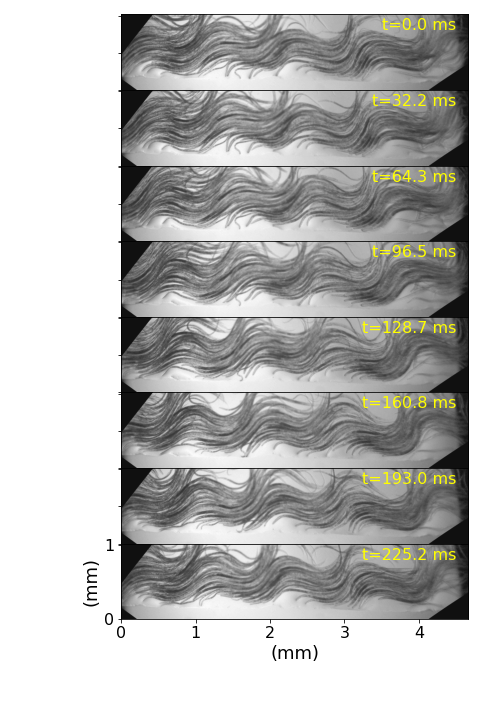}
\caption{Each panel show the same subregion of a series of frames from \eelfour.  
The edge of the drop is near the bottom of each panel.
The time of each frame from the beginning of the sequence is shown in yellow on the top right of each panel.
The $x$ and $y$ axes are in mm.
\label{fig:eels4}}
\end{figure}

\begin{figure} 
\includegraphics[width=3.5in, trim=5 5 0 0, clip]{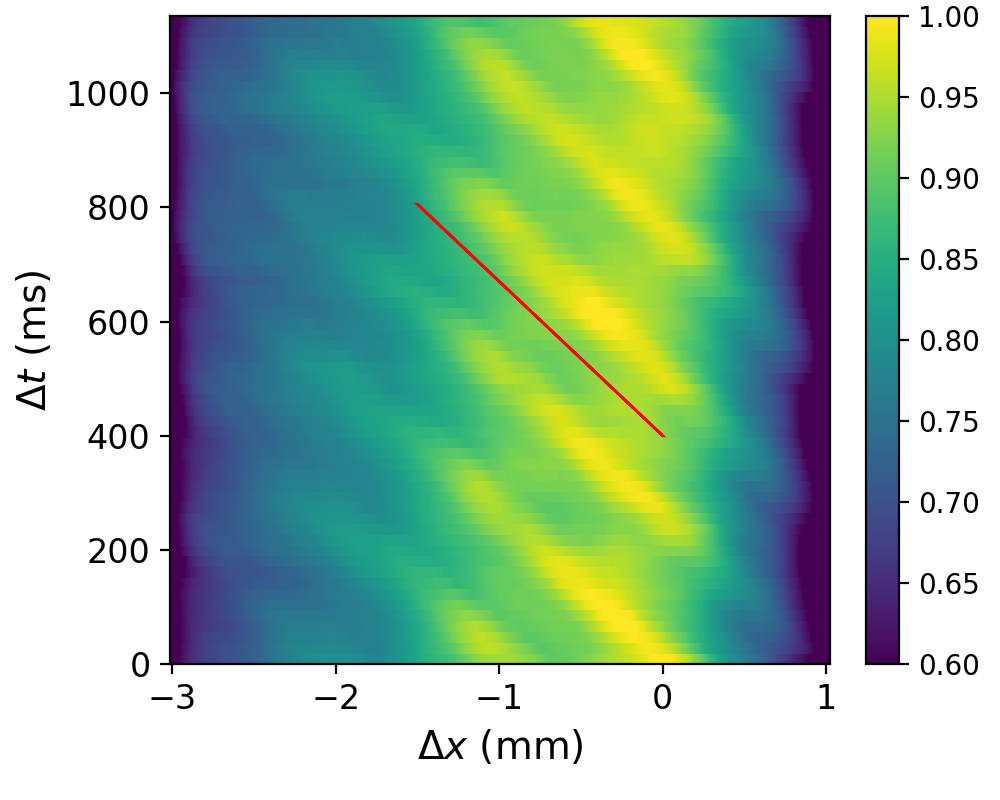}
\caption{Correlation function computed 
using equation 
\ref{eqn:carr} from image intensity  as a function of spatial shift $\Delta x$ and time delay
$\Delta t$.  
The metachronal wave speed depends on the slope of the ridges.
The estimated metachronal wave speed of $v_{\rm MW}  = 3.7$ mm/s is shown with the red segment.
\label{fig:conv}}
\end{figure}

\begin{figure} 
\includegraphics[width=3.5in, trim=5 30 10 20, clip]{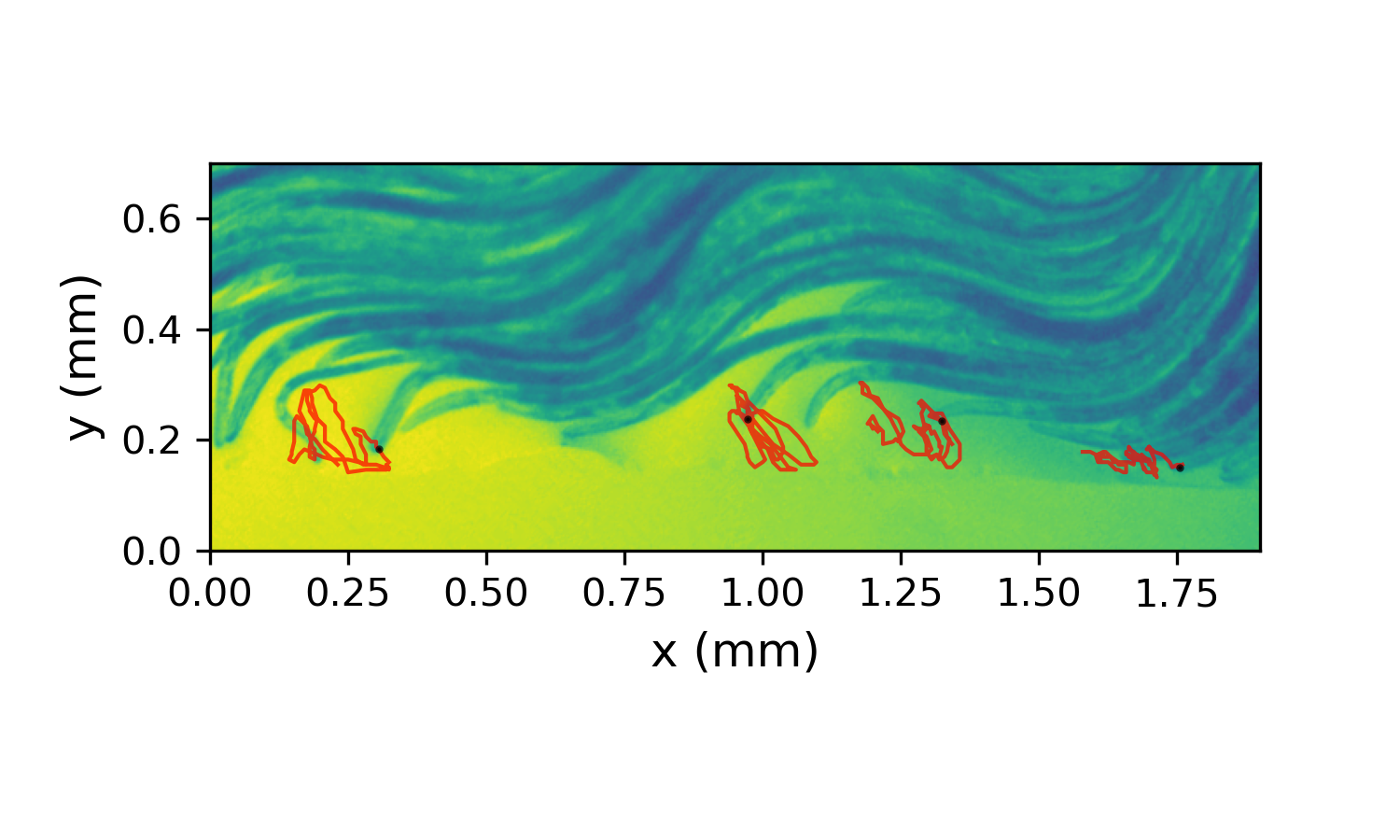}
\caption{Head positions for 4 eels were tracked over 2 seconds of
video and their trajectories are shown in red on the image.
The black dots show the location of the eel heads at the same time as the video frame.
The eels don't advance forward very quickly or at all while they are engaged in the 
metachronal wave.  The amplitude of back and forth motion is about $A_{\rm MW} \sim 0.07$ mm, and exceeds
that of the freely swimming eel.
 \label{fig:head}}
\end{figure}

\section{Observations of metachronal waves at high concentrations}
\label{sec:high}

At high concentration and 
a few minutes after the drop is placed on the slide, the eels collect near the edge of the drop, where
the air/fluid boundary touches the slide, and just within the outer rim of the drop. 
Collective motion in the form of a traveling wave becomes progressively stronger and can be seen without
magnification by eye as the vinegar eels are about 1 mm long (see Figure \ref{fig:wide}).

In Figures \ref{fig:wide}a and \ref{fig:eels4} we show frames from taken from \eelfour. 
The frames in Figure  \ref{fig:eels4}  have been rotated to orient the drop edge horizontally and at the bottom
of each panel.  To aid in comparing the frames at different times, we geometrically distorted each frame
with a near identity quadratic coordinate transformation so as to make the boundary horizontal.
The transformation used is $(x,y) \to \left( x, y- \frac{1}{2R_c} (x - x_c)^2 \right) $ with $x_c$ the $x$ coordinate
of the center of the image and $R_c$ is a radius of curvature.
Due to surface tension the actual drop edge is curved,  with a radius of curvature
of about $R_c \approx 7$ mm.  

Using frames from the rotated and distorted video 
we created a time series of one dimensional
arrays by integrating intensity along the vertical axis of the image.   
The vertical distance integrated is 1 mm and covers the frames in the series shown in Figure \ref{fig:eels4}.
This integration gives an intensity array
$\rho(x,t)$ as a function of time $t$ with 
$x$ axis parallel to the drop edge.   
We use $\rho(x,t)$ to estimate the metachronal travel
speed.   We compute a correlation function, shown in Figure \ref{fig:conv},
\begin{equation}
C(\Delta x, \Delta t) = \frac{\int dx\ \rho(x,t) \rho(x + \Delta x,t + \Delta t)}{\int dx \ \rho(x,t)^2  } \label{eqn:carr}.
\end{equation}
where $\Delta x$ is a horizontal shift and $\Delta t$ is a time delay.
The ridges in Figure \ref{fig:conv} are regions of higher intensity that propagate as a wave and their slope,
shown with a red segment, 
is the metachronal wave speed, $v_{\rm MW}$.  
We estimate the metachronal wave speed by shearing the correlation function image 
until the ridges are vertical.  
The uncertainty in $v_{\rm MW}$ is estimated from the range of shear values 
that give vertical ridges upon visual inspection
of the sheared correlation array.
We estimate the metachronal wavelength $\lambda_{\rm MW}$ with a Fourier transform of the 
orientation angles array shown in Figure \ref{fig:quiver} (which is discussed in more detail below).  
The size of the error is based on the estimated covariance of a Gaussian fit to 
the Fourier transform. We checked that this wavelength was consistent
with that measured from the distance between peaks in the correlation function shown in Figure \ref{fig:conv}.
The wavelength and wave speed also give a metachronal wave oscillation frequency 
$f_{\rm MW} = v_{\rm MW}/\lambda_{\rm MW}$.
The measurements of the metachronal wave, $v_{\rm MW}, \lambda_{\rm MW}$, and $f_{\rm MW}$,  
are listed in Table \ref{tab:meta}.


Head positions for 4 eels were tracked by clicking on their head positions in two hundred
frames spanning  2 seconds  from \eelfour  \ 
and their trajectories are shown in red in Figure \ref{fig:head}.
The eels don't swim forward very quickly. 
The four eels were chosen because their heads were easiest to identify
during the 2 s video clip.    The amplitude of back and forth motion for the eel heads
is about $A_{\rm MW} \sim 0.07$ mm.
This amplitude is an estimate for the amplitude of motion
for eels engaged in the metachronal wave and it exceeds the amplitude of motion $A_u \sim 0.045 $ mm
in the 1 mm long freely swimming eel. 

By counting eel widths, we estimate that
$N_{\rm MW} = 13$ to 15 eels per metachronal wavelength $\lambda_{\rm MW}$ 
are involved in the traveling wave.   
However only about 8 eels per mm have heads visible near the edge of the drop.
Some of the eel heads are more distant from the edge of the drop
and are confined between other eel bodies.
For deeper water/vinegar drops, 
the number of eels per unit length in the metachronal wave is sensitive to
wetting angle \citep{Peshkov_2021}.  

The metachronal wave frequency $f_{\rm MW} \sim 4 \pm 0.2 $ Hz is significantly lower than the undulation frequency of individual freely swimming eels, $f_u \approx 6 $ Hz. 
Studies of metachronal wave formation in cilia and flagellate bacteria have found that
as the filaments or flagella enter a traveling wave state, their frequency of oscillation {\it increases}
 because hydrodynamic drag on the filaments is reduced 
when they are collectively beating in a wave pattern \citep{Niedermayer_2008,Brumley_2012}.
However, here we find that the metachronal wave frequency is lower than
that of the freely swimming eels.   Since eels swimming along the edge of the 
drop do not exhibit a lower undulation frequency, the reduced frequency must
be due to interactions between organisms and 
we infer that interactions between neighboring eels reduce, rather than increase, their 
oscillation frequency. 

\subsection{Body orientations}
\label{sec:body_orient}

Figure \ref{fig:eels4} suggests that when engaged in the metachronal wave, portions of the eel's bodies spend more
time at some orientation angles than others. 
Figure \ref{fig:head} shows that during some phases of the wave, the eel heads move away from their neighbors.
There are larger gaps between eels at some phases of the wave. 
These observations suggest there are deviations from sinusoidal motion. 
In this section we measure body orientations from the video frames to quantitatively examine this possibility.

\begin{figure*}[htbp] 
\includegraphics[width=7.5in, trim=10 0 10 0, clip]{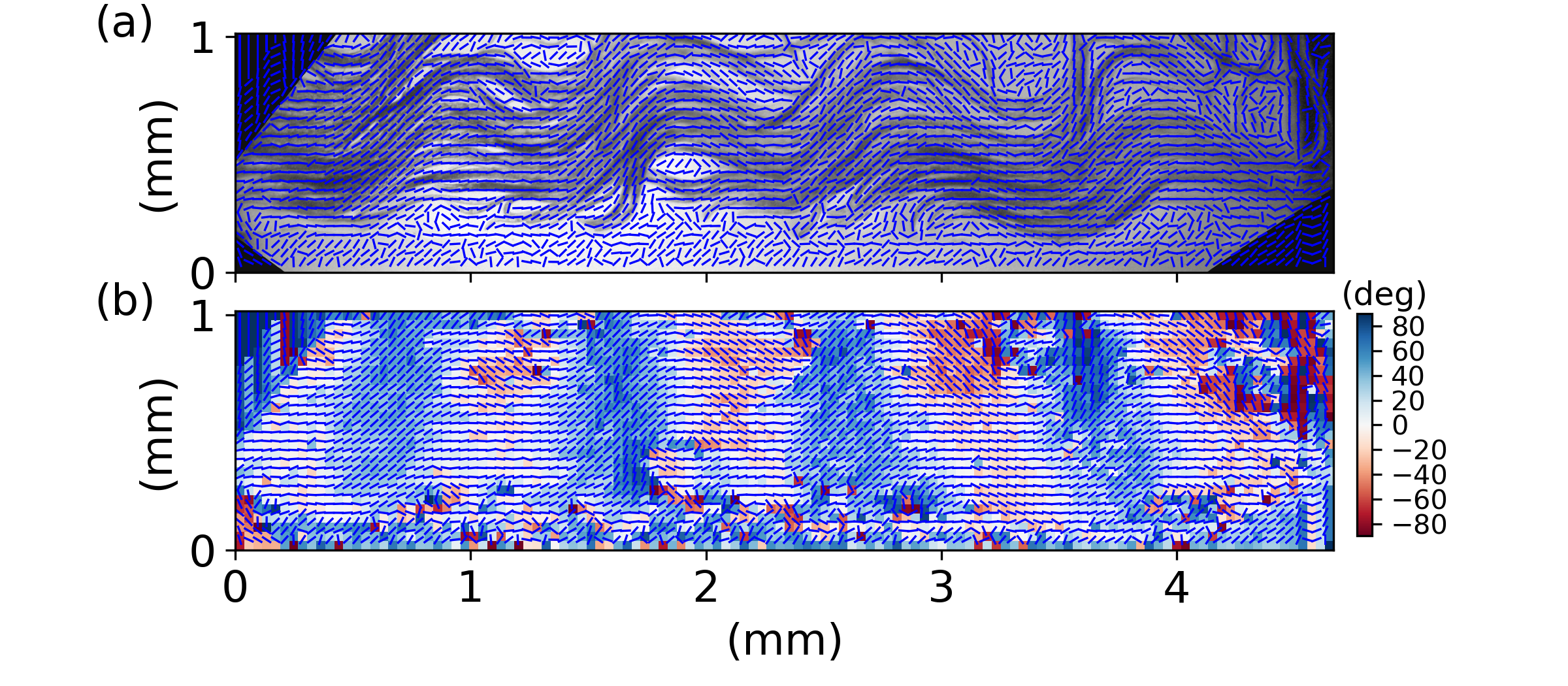}
\caption{Body orientation angles.   In (a) and (b) panels the blue segments are oriented 
with  the 
means of locally computed histograms of oriented
gradients.  The histograms of oriented gradients were 
computed from one of the images in Figure \ref{fig:eels4} from \eelfour.
The same image is shown in gray-scale in panel (a).
The color image in panel (b) displays the orientation angles, with
color-bar on the right in degrees.  
}
\label{fig:quiver}
\end{figure*}

To measure the local orientation of the eel bodies we compute local 
histograms of oriented gradients (HOG). 
These histograms are commonly used in object recognition software \citep{Dalal_2005}.
Figure \ref{fig:quiver} was made from one of the panels shown in Figure \ref{fig:eels4}.
In each 12x12 pixel square cell in the image, we computed histograms of oriented gradients
with the \texttt{hog} routine that is part of the image processing python package \texttt{scikit-image}.
We use unsigned gradients so orientation angles lie between $[-\pi/2, \pi/2]$. 
At each cell an average direction was computed using the histograms and these
are plotted as blue segments on top of the original video frame in  Figure \ref{fig:quiver}a.
In  Figure \ref{fig:quiver}b, the same blue segments are plotted on top
of a color image with color showing the angles themselves.  The color bar on the right
relates orientation angle to color, with white corresponding to a horizontal orientation.
In non-empty regions,
we estimate an uncertainty less than $\pm 20^\circ$ in the orientational angles 
based on inspection of Figure \ref{fig:quiver}a.  

\begin{figure} 
\includegraphics[width=3.5in, trim=5 25 0 35, clip]{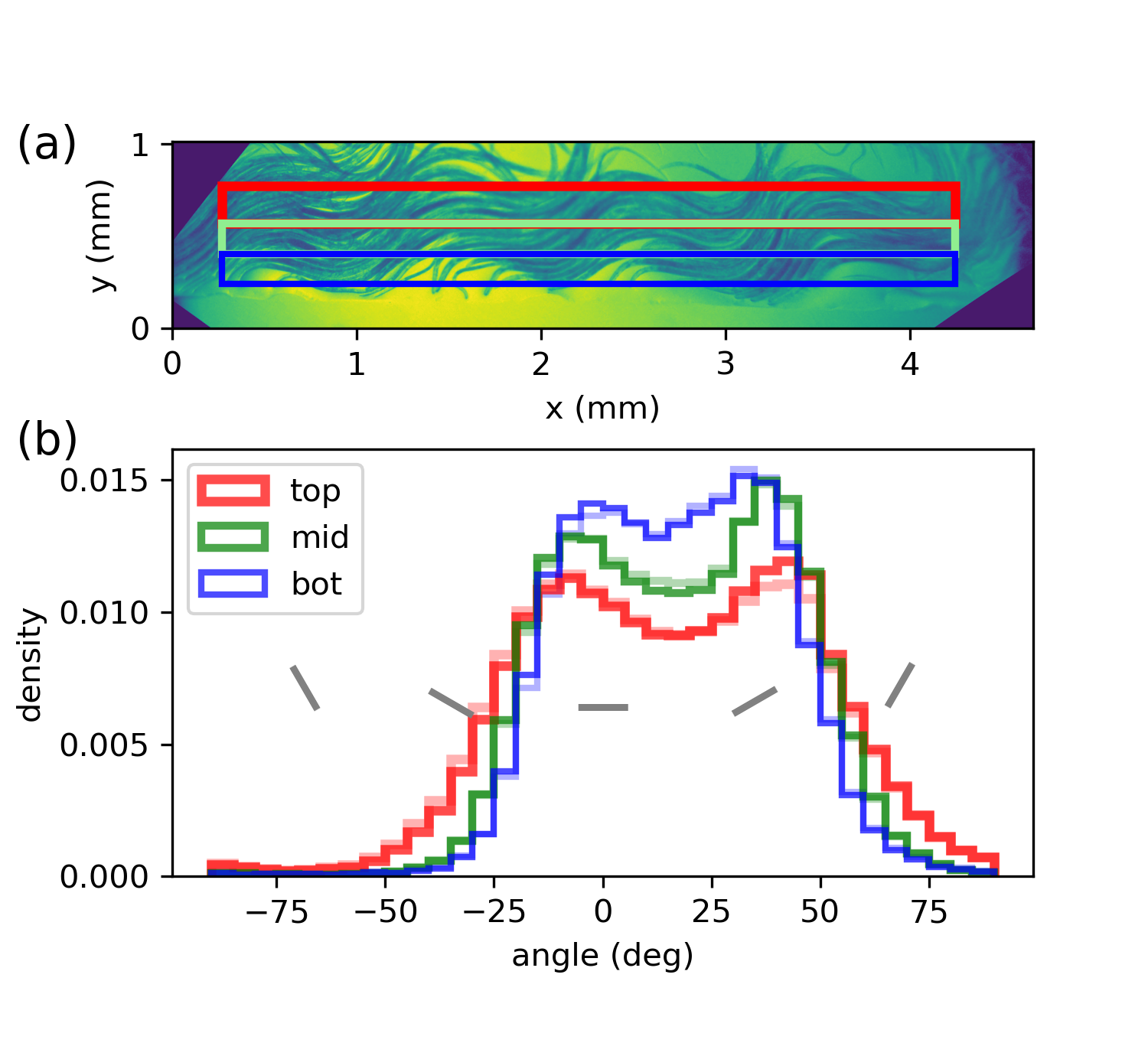}
\caption{Distributions of orientation angles in the metachronal wave.
(a) Three rectangular regions are shown on top of one of the image frames.
The color and line width showing each region is the same as in panel (b).
(b) Normalized distributions of orientation angles in the three rectangular image regions.
The histograms were computed using orientations like those shown in Figure \ref{fig:quiver}, but using 
 200 video frames from \eelfour \ spanning a duration of 2 s.   
The higher opacity lines are histograms computed with intensity weights with
regions of higher eel density contributing more to the histogram.  
The lighter lines are histograms computed without weighting. 
The distribution is narrower near the edge of the drop.
The gray bars have orientation equivalent to their $x$ coordinates on the plot and are
plotted at multiples of $30^\circ$.
The difference in the two peak heights in each distribution suggest that there are 
deviations from sinusoidal shapes and motions.  
\label{fig:hist}}
\end{figure}

To examine statistical variations in the body orientations  
 we computed distributions from the orientation angles 
(like those shown in Figure \ref{fig:quiver})
but using 200 video frames from \eelfour \ spanning a duration of 2 s.
A large number of video frames were used to average over the different
phases of the wave.    Orientation angle distributions are shown in Figure \ref{fig:hist}b.

Three rectangular regions are drawn in Figure \ref{fig:hist}a
on one of the image frames and each region is plotted with the same color and thickness
 line as used in Figure \ref{fig:hist}b. 
In  Figure \ref{fig:hist}b  we 
show distributions of orientation angles measured in these three rectangular regions.
The three region centers have different distances 
from the edge of the drop, $0.47, 0.29$ and $0.13$ mm. 
The higher color opacity lines in Figure \ref{fig:hist}b are distributions computed with weights 
 so that regions of high eel intensity contribute more to the histogram. 
The lighter and lower opacity lines are distributions computed without weighting. 
The difference between the higher and lower opacity lines shows that the orientation angle distributions 
are not  sensitive  
to local variations in image intensity.
The red rectangular region (plotted with wider lines) is more distant from the edge of the drop 
than the blue region.  
The red histogram is wider than the blue one, indicating
that there is a wider range of body orientation angles more distant from the drop edge.

The distributions shown in Figure \ref{fig:hist}b have a trough and are asymmetric or lopsided, with
one peak higher than the other.   This asymmetry is not expected as 
a sine wave has distribution of orientations (computed from its slope) that would
be symmetrical about a mean value.  
Models for the orientation angle distribution are discussed further in section \ref{sec:orient}. 

In summary, we find that for vinegar eels engaged in a metachronal wave, the distribution 
of body orientation angles has two peaks of different heights and depends on distance to the drop edge.
The asymmetry in the orientation angle distribution and inspection of eel heads near the drop edge implies that eel body shapes and motions are not perfectly sinusoidal.  
This contrasts with our study of the freely swimming eels in section 
\ref{sec:low} where we found 
that the shape and motion of freely swimming eels is nearly sinusoidal.

\section{Oscillator models for traveling waves}
\label{sec:osc}

Experimental observations have shown that motility of swimming nematodes, such as {\it C. elegans},
is due to the propagation of bending waves along the nematode's body length \citep{Gray_1964};
(for a summary of nematode locomotion neurobiology, see \citep{Cohen_2014}).
The bending waves consist of alternating phases of coordinated dorsal and ventral muscle contractions
and extensions \citep{Haspel_2011}. 
During locomotion, motor neurons excite muscles on either (ventral/dorsal) side of the body while  inhibiting muscles on the opposite side. 

The gait of {\it C. elegans} adapts to the mechanical load imposed by the environment \citep{Boyle_2012}.
Swimming involves higher frequency and longer
wavelength undulations than crawling on agar, though both behaviors may be part of a continuous
spectrum of neural control \citep{Niebur_1991,Lebois_2012}.    
Oscillation frequencies also decrease for {\it C. elegans} 
swimming in higher viscosity aqueous media \citep{Sznitman_2010}.
Proprioception is when sensory receptors in muscles or other tissues 
are sensitive to the motion or position of the body. 
In models for nematode locomotion, the sensitivity to environment
involves proprioceptive integration or feedback on the neuronal control model 
\citep{Niebur_1991,Berri_2009,Wen_2012,Cohen_2014}. 

Experiments of restrained {\it C. elegans} \citep{Wen_2012} show that the bending
 of the posterior regions requires anterior bending (see Figure 3 by \citet{Wen_2012}).
 If the nematode is held fixed at its middle, the body can undulate 
 between head and constraint, but past the constraint to the tail, there
 will be no undulation.
These experiments suggest that the body itself lacks central pattern generating 
circuits and motivates locomotion models that rely on an oscillator in the head \citep{Wen_2012}.

To create a model for collective motion in the vinegar eels,
we assume that the waves that propagate down the nematode's body are initiated at the organism's head. 
We use the phase of the head's back and forth motion with respect to its mean position
to describe the state of each organism
and we model our ensemble of eels as a chain of phase oscillators.
In the absence of interactions, each oscillator has 
 intrinsic frequency equal to the oscillation frequency of a freely swimming eel.
 Because the mean positions (averaged over the oscillation period) 
 of the eel's heads drift very slowly (see Figure \ref{fig:head}),  
 we neglect drift in the mean or averaged (over a period) oscillator positions.
 Here the oscillator phase is associated with back and forth motion of an eel head because the head
 is assumed to be the source of the body wave.   This differs from the models 
 by \citet{Niedermayer_2008,Brumley_2012} where the phase describes motions of a cilium or flagellum tip.

When the vinegar eels are engaged in metachronal waves, the organisms are often touching each other.
\citet{Chelakkot_2021} simulated steric interactions between active and elastic filaments in arrays
and found that short-ranged steric inter-filament interactions can account for 
formation of collective patterns such as metachronal waves.
Because undulation frequency of {\it C. elegans} is slower when 
under mechanical load imposed by the environment, 
we assume that steric interactions in our vinegar eels 
reduce the phase velocity of oscillation.

To construct a model for metachronal waves, we consider the head of a single organism
to be an oscillator and we consider ensembles of $N$ oscillators.  
The $i$-th oscillator can be described with a phase $\theta_i$ 
and a frequency of oscillation or a phase velocity 
$\frac{d\theta_i}{dt} =\dot \theta_i $.
Here $i$ is an integer index and $ \theta_i$ is a function of time $t$.

Collective phenomena involving synchronization of oscillators
has been described with different nomenclature. 
Following \cite{Acebron_2005,Niedermayer_2008}, 
 a {\it synchronized } state of an ensemble of $N$ oscillators is one where all oscillators 
have identical phases, $\theta_i(t) = \theta_j(t)$ for all $i,j \in (0, 1, ... N-1) $.
A {\it phase-locked } or {\it frequency synchronized} state \citep{Ermentrout_1986,Ermentrout_1990,Ren_2000} 
is one where all oscillators have 
  identical phase velocities $\dot \theta_i(t) = \dot \theta_j(t)$ for all $i,j  \in (0, 1, ... N-1)$.
An {\it entrained} state has identical mean phase velocities 
 $\tilde \omega_i =\tilde \omega_j$
  for all $i,j \in (0, 1, ... N-1) $.   The time average of 
 the phase velocity can be computed with an integral
 over time, 
 $\tilde \omega_i   = \lim_{t \to \infty} \frac{1}{t} \int_0^t \dot \theta (t) dt $,  
 or by integrating over an oscillation period if oscillator motions become periodic.

For a chain of oscillators, the index $i$ specifies the order in the chain.
One type of traveling wave is a non-synchronous phase-locked state characterized by
a constant phase delay or offset between consecutive oscillators in a chain or loop of oscillators.
In other words 
$\theta_{i+1} = \theta_i + \chi$  for consecutive oscillators,  where $\chi$ is the phase delay
and $\dot \theta_i \ne 0$ for all $i$. 
If individual oscillators undergo similar periodic motions, then another type of traveling wave is a non-synchronous
but entrained state characterized by a time delay between the motions of consecutive oscillators.
In other words $\theta_i(t + \tau) = \theta_{i+1}(t)$ with time delay $\tau$.
In this case the phase velocities would be periodic and need not be constant.
Both types of traveling waves involve periodic oscillator motions and are known in the biological literature 
as metachronal waves.

\subsection{Local Kuramoto models}

The Kuramoto model \citep{Kuramoto_1975,Kuramoto_1987,Acebron_2005}
consists of $N$ oscillators, that mutually interact via a sinusoidal interaction term 
\begin{equation}
\frac{d \theta_i}{dt} = \omega_i +  \sum_{j=1}^N K_{ij} \sin (\theta_j - \theta_i) 
\end{equation}
where $K_{ij}$ are non-negative coefficients giving the strength of the interaction
between a pair of oscillators.    In the absence of interaction, the $i$-th oscillator would have a 
constant phase velocity $\omega_i$ which is called its intrinsic frequency.

With only nearest neighbor interactions a well studied model,  
sometimes called a local Kuramoto model, is described by
\begin{equation}
\frac{d \theta_i}{dt} = \omega_i + K \left[ \sin(\theta_{i+1} - \theta_i) + \sin (\theta_{i-1} - \theta_i) \right]
\label{eqn:local_model}
\end{equation}
\citep{Ermentrout_1986,Ermentrout_1990,Ren_2000,Muruganandam_2008,Tilles_2011,Denes_2019}.
At low values of positive interaction parameter $K$, the oscillators are not affected by their neighbors.
At higher $K$, the oscillators cluster in phase velocity, and the number of clusters 
decreases until they fuse into a single cluster that spans the system.  
At and above a critical value of $K=K_s$ the entire system must enter a global phase-locked state \citep{Aeyels_2004}.
Above the critical value $K>K_s$, there can be multiple stable phase-locked attractors, each with its
own value of global rotation rate $\Omega = \frac{1}{N} \sum_i \omega_i$  \citep{Zheng_1998,Tilles_2011}.


What fraction of possible initial conditions would converge onto a phase-locked solution that is not synchronous?   
 The set of initial conditions that converge onto a particular
solution are called its {\it basin of attraction}.
The basins of attraction for traveling wave solutions (or non-synchronous phase-locked states)  
are smaller than that of the synchronous
state \citep{Tilles_2011,Denes_2019}.  Using random and uniformly generated initial phases in $0$ to $2 \pi$
 for each oscillator, the system is more likely to enter a synchronous rather 
than a traveling wave state.  

Because well studied local Kuramoto models like that of equation \ref{eqn:local_model} are more
likely to enter a synchronous than  a traveling wave state,
they do not capture the behavior illustrated by our vinegar eels, or other systems
that exhibit metachronal waves such as chains of cilia \citep{Niedermayer_2008} or 
flagella on the surface of {\it Volvox carteri} alga colonies  \citep{Brumley_2012}.
Relevant models should exhibit a larger basin of attraction
for traveling wave states than for the synchronous state.


In models for metachronal waves in cilia or flagellates 
 \citep{Vilfan_2006,Niedermayer_2008,Brumley_2012}
the end of a filament moves in a plane and on
a trajectory of radius $R$ from a central position with phase $\theta$ in polar coordinates.  
Active forces are induced via tangential forces exerted on the filament.
Interactions between the oscillators are based on hydrodynamic interactions 
between pairs of filaments and are computed using Stokes equation which is valid at low Reynolds number 
\citep{Vilfan_2006,Niedermayer_2008,Uchida_2011,Brumley_2012}.
Motion is over-damped so the equations of motion are a balance between driving
and hydrodynamic forces.  The filament velocities are computed as a function of their 
positions and it is not necessary to compute accelerations. 
The equations of motion describe motions of the phase,  radius and orientation angle 
of the end of the filament's trajectory. However, if the distance between
filaments is large compared to the radius of motion,
 the dynamical system can be approximated with nearest neighbor interactions
and neglecting variations in the radius or plane of motion \citep{Niedermayer_2008}.
This gives a local oscillator chain model dependent only on phases.

\subsection{An oscillator model based on heads that overlap}
\label{sec:omodel}

We desire a model that has a wide basin of attraction for traveling wave states, similar
to those by \citet{Niedermayer_2008,Brumley_2012}.  The oscillator chain model by 
 \citet{Niedermayer_2008} included sine and cosine terms of the sums and differences
 of pairs of phases  and that by  \citet{Brumley_2012} included both radial and phase motions.
We can similarly assume
that motion is over-damped and can be described
by equations for phase and phase velocity and lacking phase accelerations.  
Since steric interactions
are likely to be important, we can adopt a model with only nearest neighbor interactions, 
as did \citet{Niedermayer_2008}.  However, opposite to
the hydrodynamic interaction models, the interactions between our
eels are likely to be strong,  and they should reduce the oscillator 
phase velocity rather than increase it.
We observe that eel heads near the edge of the drop
(see Figures \ref{fig:eels4}, \ref{fig:head}) were not near other eel bodies during  
portions of the traveling wave.  If undulation is generated at the eel head, then
interactions on it are only strong during about half of the head's oscillation cycle. 

\begin{figure}
\centering
\includegraphics[width=3.0in, trim=0 0 0 0, clip]{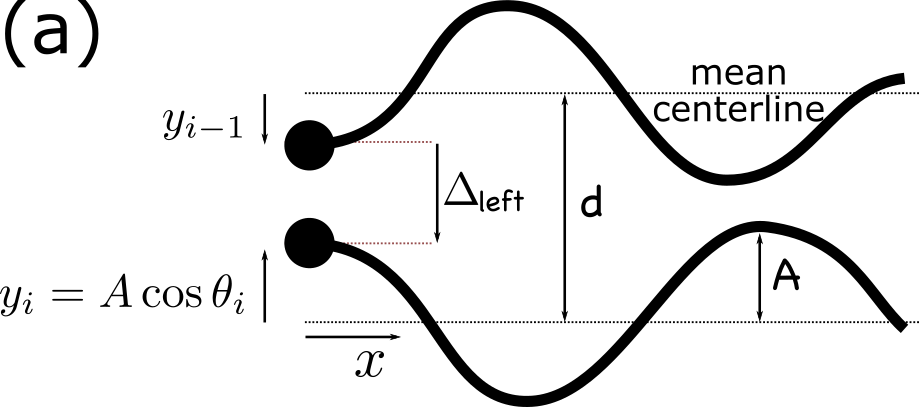}
\includegraphics[width=2.7in, trim=0 0 0 0, clip]{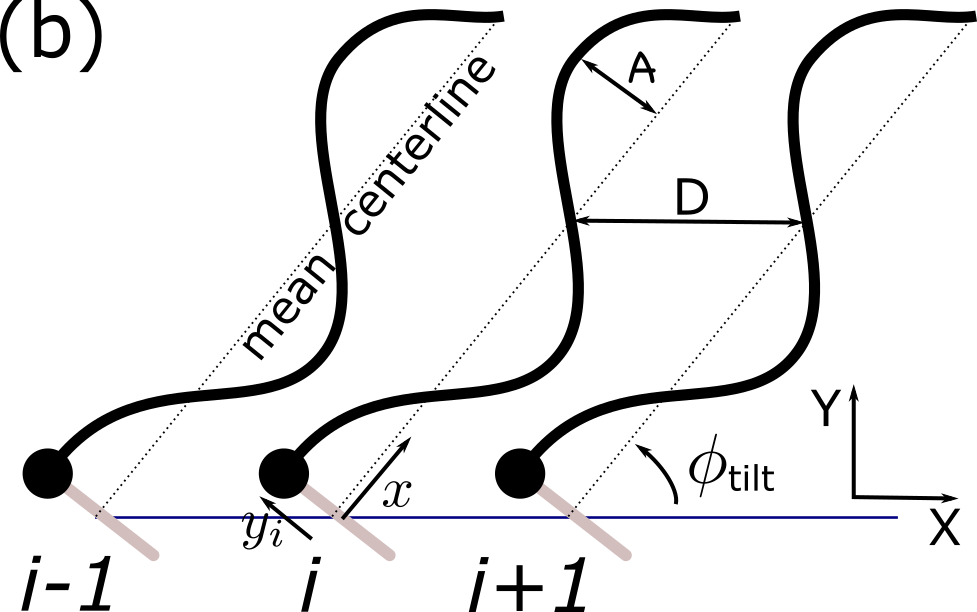}
\caption{ 
(a) Two eels undulate with amplitude $A$
but without moving their mean centerlines.  The two mean centerlines are  
 shown with dotted lines and are separated by distance $d$. 
The eal heads are shown with large black dots.
We assume that the undulation on the body is initiated by oscillators in the eel's heads.
The oscillators have 
 phases $\theta_i$ and $\theta_{i-1}$.  
When $\Delta_{\rm left} = d + A \cos \theta_{i} - A \cos \theta_{i-1}<0$,  the eel heads overlap and
 steric interaction would slow their motion.   
(b) Three consecutive eels are tilted by angle $\phi_{\rm tilt}$ with 
respect to the horizontal direction.  
The oscillator in the $i$-th eel's head is more strongly influenced by the motions of
the organism to its left (with index $i-1$) than the one to its right
(with index $i+1$).   At lower tilt angle $\phi_{\rm tilt}$,  the interactions are increasingly lopsided. 
\label{fig:overlap}
}
\end{figure}

Consider two eels oriented horizontally as shown Figure \ref{fig:overlap}a with $x$ the horizontal axis 
and $y$ the vertical one.
The eels undulate with amplitude $A$ and without varying the head's $x$ position or 
the orientation of its mean centerline, which is shown with dotted lines.
The $y$ position of the $i$-th head 
\begin{equation}
y_i = A\cos \theta_i -id,
\end{equation}
where $d$
is the distance between the neighboring eel's mean centerlines.
The phase of oscillation is given by the angle $\theta_i$.
The distance between the two heads with index $i$ and $i-1$ is 
\begin{equation}
\Delta_{\rm left} = d + A \cos \theta_{i} - A \cos \theta_{i-1}.\end{equation}
 The eels with index $i$ and $i-1$ overlap near their heads if the left-sided overlap function 
\begin{equation}
o_{\rm left}(\theta_{i-1},\theta_i) = \frac{\Delta_{\rm left}}{A}=  \cos \theta_i - \cos \theta_{i-1} + \beta < 0, \label{eqn:o_left}
\end{equation}
where the dimensionless overlap parameter 
\begin{equation}
 \beta \equiv \frac{d}{A}.  \label{eqn:beta}
\end{equation}
We assume that 
a strong steric interaction on the $i$-th eel's head would reduce its phase velocity 
when $o_{{\rm left}}(\theta_{i-1},\theta_i)<0$.
Otherwise, the eel head's phase velocity would remain at its intrinsic phase velocity.
Because the eels tend to be closer together than the amplitude of undulation  when they are involved
in a metachronal wave, 
we expect $\beta$ to be smaller than 1. 
The amplitude $A$ of body motions for eels engaged in the metachronal wave need not be the same as that of the freely swimming eel, $A_u$.

Consider three eels oriented at an angle as shown in Figure \ref{fig:overlap}b.
The oscillator in the $i$-th eel's head is more strongly influenced by the motions of
the organism to its left (with index $i-1$) and less so by the one to its right
(with index $i+1$).   When the eels are tilted with respect to the edge of the
 drop, we expect directed interactions where the phase of the eel's head is
primarily influenced by its nearest neighbor on one side.

A modification to the local Kuramoto model with directed or one-sided nearest neighbor interactions 
\begin{equation}
\frac{d \theta_i}{dt} \omega_0^{-1} =  1 - K  f(\theta_{i-1},\theta_{i}). 
\end{equation}
Here positive and dimensionless parameter $K$ describes the strength of the interaction. 
The nearest neighbor interaction function $0<f(\theta_{i-1},\theta_{i})\le 1$, reduces
the phase velocity and mimics the role of one-sided steric interactions.
The intrinsic angular phase velocity $\omega_0$ is the same for each oscillator.
We work with time in units of $\omega_0^{-1}$ which is equivalent to setting $\omega_0=1$.

One choice for the interaction function should give 1 if the overlap function 
$o_{\rm left} $ (defined in equation \ref{eqn:o_left}) is negative
and there is an overlap and gives 0 otherwise.  This choice neglects eel body width.
We have checked with numerical integrations that 
a numerical model based on a Heaviside step function can robustly give traveling wave solutions. 
However, numerical integration of a discontinuous function with a conventional numerical integrator
can give results that are dependent on step size or sensitive to round-off or discretization errors. 
To mitigate this problem we use a smooth function to approximate the step function, 
$f(\theta_{i-1},\theta_{i}) = \frac{1}{2} \left[ 1-\tanh \frac{o_{\rm left}(\theta_{i-1},\theta_i)}{h_{ol}}\right]$ 
where dimensionless parameter $h_{ol} $ sets the abruptness of the transition of the function from 0 to 1.
In the limit of small $h_{ol}$ we recover the Heaviside function.
An oscillator model that uses this smooth function has equation of motion 
\begin{equation}
\frac{d \theta_i}{dt} \omega_0^{-1} =  1 - \frac{K}{2} 
\left[ \tanh \left(  \frac{\cos \theta_{i-1} - \cos \theta_i -  \beta }{ h_{ol} }\right) + 1\right] . 
\label{eqn:omodel}
\end{equation}

\begin{figure*}
\includegraphics[width=6.5in, trim=10 10 10 0, clip]{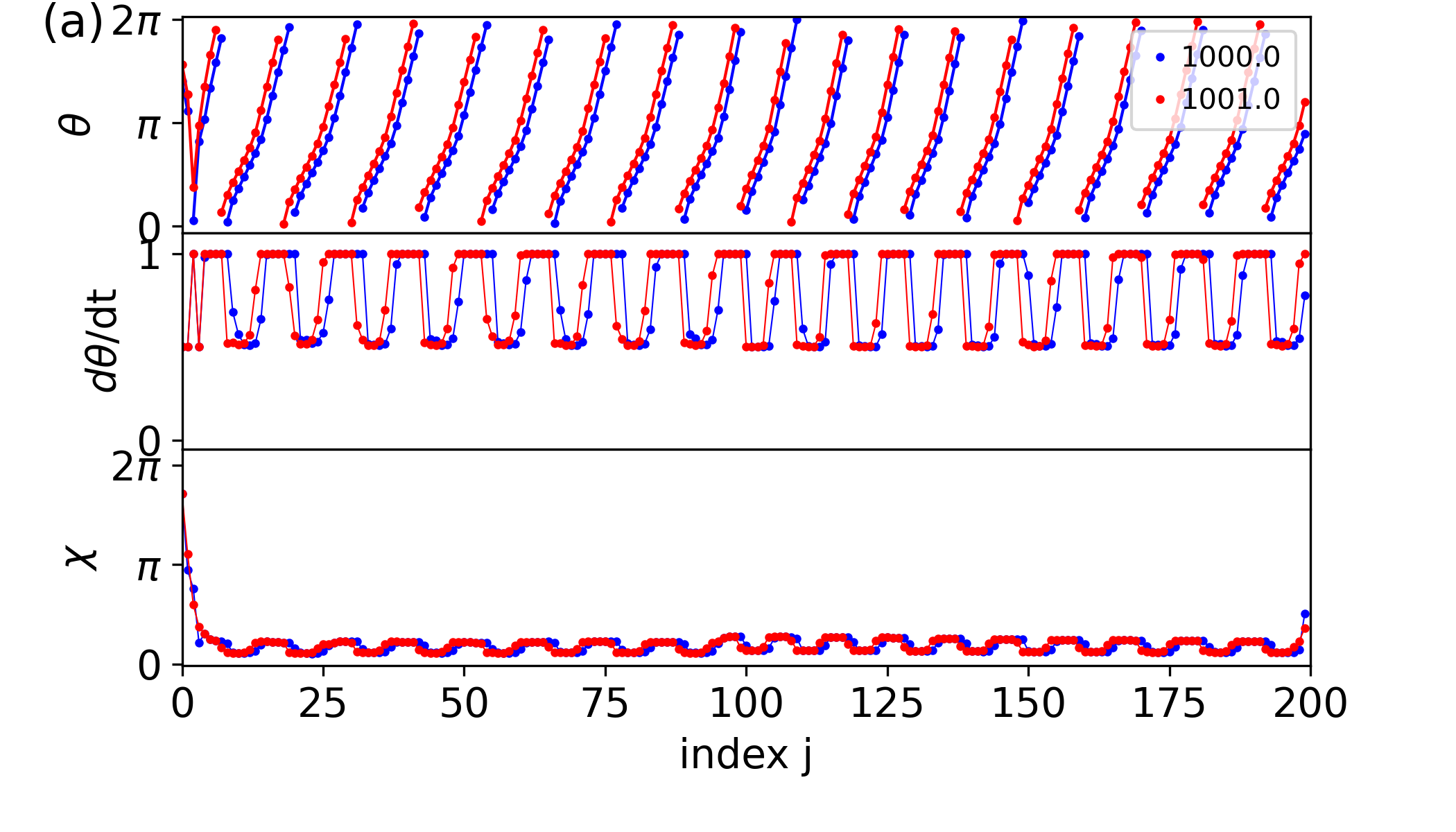}
\includegraphics[width=6.5in, trim=10 10 20 0, clip]{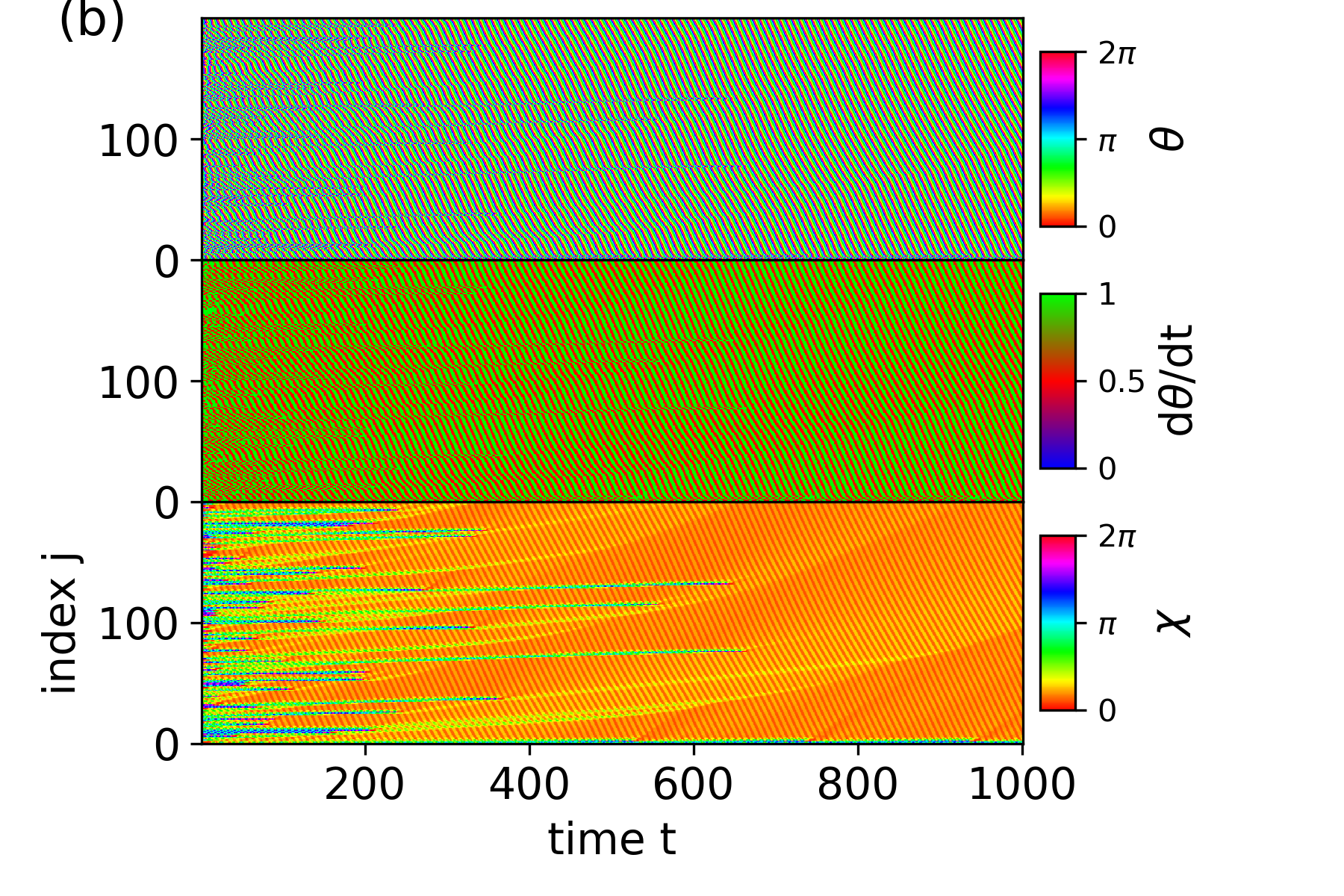}
\caption{A directed oscillator chain model numerical integration. 
Equation \ref{eqn:omodel} is integrated with  $N=200$ oscillators in a chain
with a non-periodic boundary condition and randomly chosen initial phases.   The interaction 
parameter $K=0.5$, intrinsic frequency $\omega_0=1$, overlap parameter $\beta=0.25$,
smoothness parameter  $h_{ol} = 0.05$ and 
time-step $dt=0.05$.    The system was integrated to time $t=1001$.
At the end of this integration the average phase velocity $\tilde \omega = 0.77 \omega_0$
and the average wavelength is $N_\lambda = 12$ oscillators.
(a) From top to bottom panels, the phase angles $\theta_j$, phase velocity $d \theta_j/dt$
and phase difference $\chi_j = \theta_{j+1} - \theta_j$ are plotted as a function 
of index $j$ at two different times.    
The outputs at $t=1000$ and $t=10001$ are plotted with red and blue lines.
Comparison between these two outputs shows
that they are similar but shifted by a time delay.  The system is an entrained state
which can also be described as a traveling wave state.
(b) From top to bottom panels, the images show phase angle 
$\theta_j$, phase velocity $d \theta_j/dt$
and phase difference $\chi_j $ with color shown in the color-bars on the right.
The horizontal axes is time and the vertical axes are the oscillator index $j$.
The fine diagonal features at large times are the traveling waves.
The horizontal features are discontinuities that eventually disappear as coherent regions
merge.
\label{fig:ov}
}
\end{figure*}

\begin{figure}
\includegraphics[width=3.5in, trim=15 35 0 2, clip]{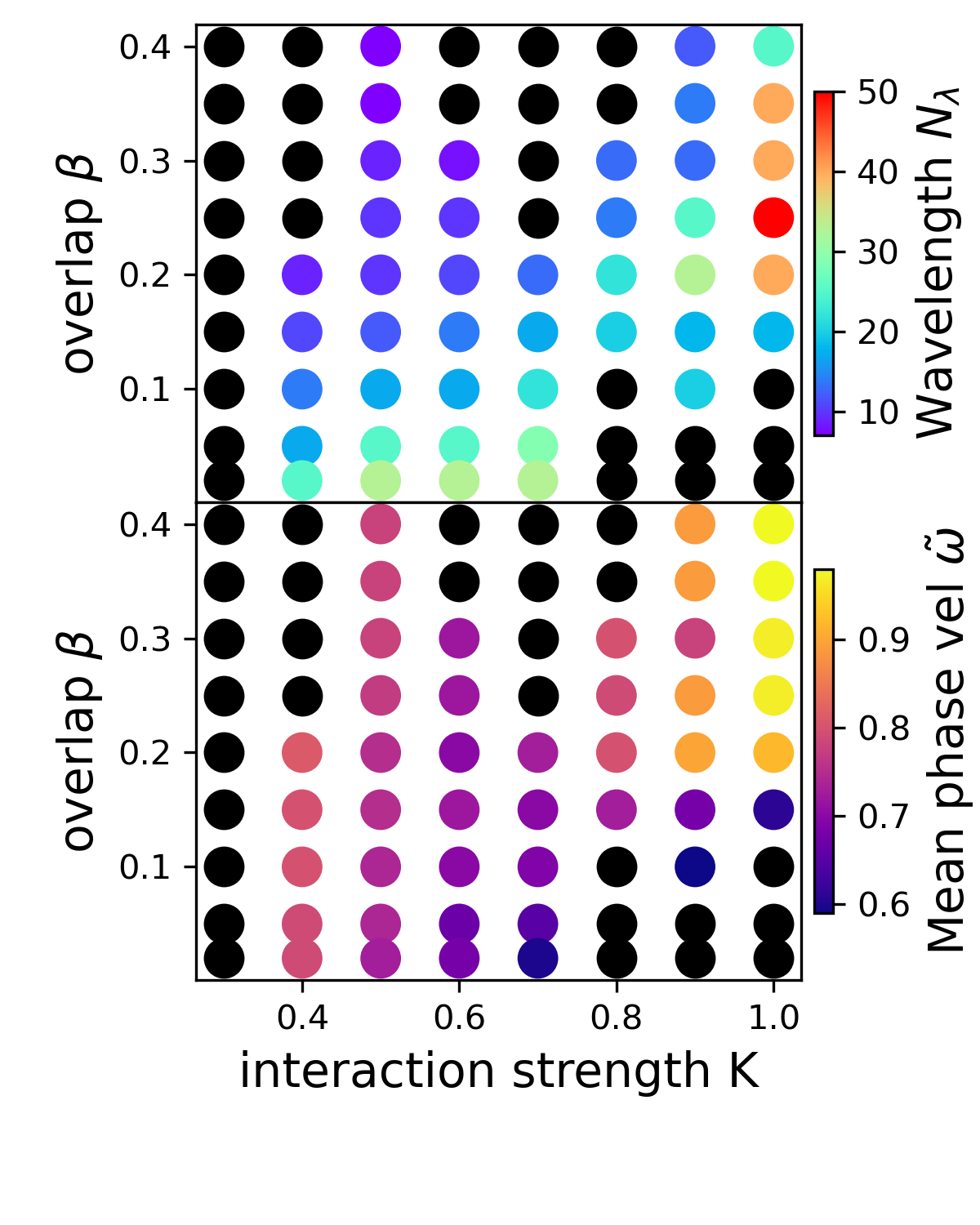}
\caption{Wavelengths $N_\lambda$ and mean phase velocity $\tilde \omega$
computed for numerical integrations at $t=1000$ of
the directed oscillator chain model given in equation \ref{eqn:omodel}.
The integrations have $N=200$ oscillators, the timestep is $dt=0.05$
the smoothness parameter is $h_{ol} = 0.1$,  the boundary is not periodic
and initial phases were randomly chosen.  
If the entire chain of oscillators did not reach a traveling wave state at $t=1000$, a black dot
is plotted otherwise the dot has color giving the wavelength $N_\lambda$ (top panel)
and mean phase velocity $\tilde \omega$ (bottom panel).
The $x$ axis is the interaction strength parameter $K$ and the $y$ axes
are the overlap parameter $\beta$.
 \label{fig:omodel}
}
\end{figure}

\subsection{Numerical integrations of a directed overlap phase oscillator chain model}
\label{sec:num}

The directed overlap phase oscillator model given by equation \ref{eqn:omodel}
depends on three positive parameters, the interaction strength $K$, an overlap parameter
$\beta$ and the parameter setting the smoothness of the interaction function $h_{ol}$.
The model is also sensitive to the number of oscillators in the chain or loop $N$,
the boundary condition and the choice of initial conditions.
We integrate this model using a first order explicit Euler method.
The initial phases for each oscillator are randomly generated using a uniform distribution spanning $[0,2 \pi]$.

In local Kuramoto models, stable solutions that are present in a loop may not
be present if one link is dissolved and the loop becomes a chain \citep{Tilles_2011,Ottino_Loffler_2016}.
To ensure that traveling waves are robustly generated in our model, we purposely 
do not chose a periodic boundary condition. 
The boundary at the end of the chain or for $\theta_{N-1}$ does not affect the dynamics 
because of the direction of the interactions.
For the left boundary (with phase $\theta_0$) we set the phase velocity $\frac{d \theta_0}{dt} = (1-K) \omega_0$.
We find that a slow left boundary is less likely to excite perturbations that propagate through
the system. 

A numerical integration with $N=200$ oscillators, intrinsic frequency $\omega_0=1$, interaction 
parameter $K=0.5$,  overlap parameter $\beta=0.25$, and smoothness parameter 
$h_{ol} = 0.05$  is shown in Figure \ref{fig:ov}.  
The time step used is $dt=0.05$ and we have checked that a smaller
step size does not significantly change the integration output.
In Figure \ref{fig:ov}a the panels 
 show phase angle $\theta_j$, phase velocity $d\theta_j/dt$ and phase shift
$\chi_j = \theta_{j+1} - \theta_j$ as a function of index $j$ for an integration at two times $t = 1000$
and $t=1001$. In Figure \ref{fig:ov}b we show the same quantities but with color arrays as a function of both
index and time.  Despite the absence of a diffusive-like interaction term (similar to
that in equation \ref{eqn:local_model}), 
the model has attracting entrained or traveling wave solutions.  A comparison between
the two outputs in the top panel shows that phases at different times can be related with a time delay.
At the beginning of the integration
clusters of entrained or nearly phase-locked groups form and later merge to give a fully
entrained or traveling wave state.   This type of behavior was previously seen in the oscillator models 
developed for hydrodynamic interactions between cilia and flagella \citep{Niedermayer_2008,Brumley_2012}.

When initial conditions are random, there are initially groups of neighboring oscillators with
large phase differences and these large differences 
can remain on the same group of oscillators for many oscillation periods.
These are nearly horizontal streaks seen in the bottom panel showing phase difference 
$\chi$ in 
Figure \ref{fig:ov}b.  Had we added a diffusive-like term to our model, small wavelength
perturbations would be more rapidly damped, but such a term would also affect
the velocity and wavelength of traveling wave states.   

We ran the integration to a maximum time $t =1001$ with $\omega_0=1$
 corresponding to $1001/(2\pi)\approx 160$ oscillation periods ($2 \pi/\omega_0$).
For an oscillation frequency  of $f_u  \sim 6 $ Hz (as we observed
for our vinegar eels) this duration corresponds to 27 seconds.
The metachronal waves take a few minutes appear after the drop is placed on the slide.
The time it takes for all entrained clusters to merge in the numerical model is shorter than  
the few minutes it takes 
for traveling waves to form on a large portion of the drop edge in our concentrated
eel experiments.  However, our model is of a fixed chain of oscillators so 
it does not take into account
the time it takes for the vinegar eels to collect on the boundary or sources of noise in the system. 

At the end of the numerical integration shown in Figure \ref{fig:ov}, the average phase velocity 
$\tilde \omega = 0.77 \omega_0$  (computed from all 
oscillators at that time), the average wavelength is $N_\lambda = 12$ oscillators.   
The phase delay for the entrained state $\tau = \frac{2 \pi}{\tilde \omega N_\lambda} =  0.68$. 
The number of oscillators for a change of $2 \pi$ in phase,
$N_\lambda$,  is comparable to that we
estimated for the metachronal wave in the vinegar eels  (see Table \ref{tab:meta}).
The average phase velocity ratio $\tilde \omega/\omega_0 $ is near but somewhat higher than  
the ratio of metachronal wave to freely swimming undulation frequency   
$f_{\rm MW}/f_u \sim 0.67$ that we estimated for the vinegar eels (listed
in Table \ref{tab:meta} and discussed in section \ref{sec:high}).

If all phases are initially set to the same value, the dynamical system described by equation 
\ref{eqn:omodel} remains in a synchronous 
state. However, if some noise is introduced into the system (in the form 
of small stochastic perturbations on each oscillator) then the system is likely to enter the traveling
wave state even with flat initial conditions.  
The basin of attraction for the traveling wave
state is significantly larger than that of the synchronous state.

With a fixed value of smoothness parameter $h_{ol}$, we integrated equation
equation \ref{eqn:omodel} for different values of interaction parameter $K$ and overlap
parameter $\beta$.   These integrations have random initial conditions and non-periodic boundary,
as described above, intrinsic frequency $\omega_0=1$ and smoothness parameter $h_{ol} = 0.1$. 
At $t=1000$ we inspected plots like those in Figure \ref{fig:ov} to see if
the system was in an entrained state.  If so, we measured the mean wavelength $N_\lambda$
and the mean phase velocity $\tilde \omega$.
In Figure \ref{fig:omodel}
points are plotted as a function of $\beta$ and $K$ and with color set by their wavelength $N_\lambda$ (top panel)
or mean phase velocity $\tilde \omega$ (bottom panel).
Systems that exhibited discontinuities at the end of the simulation (other than at the left boundary)
are plotted in black.  
A fairly wide range of interaction and overlap parameters robustly gives entrained 
or traveling wave states.   

At larger overlap parameter, $\beta$, the oscillators spend less time overlapped 
 and this tends to give a shorter wavelength and higher mean phase velocity $\tilde \omega$
 in the entrained state.
If eels are more distant from each other or have lower amplitude oscillations then $\beta$ is larger. 
At large overlap parameters $\beta \gtrsim 0.4$ (on the top of each panel in Figure \ref{fig:omodel}) 
the system is less likely to be in a traveling wave state at $t=1000$.   
This is due to clusters of oscillators that begin with large phase differences between neighbors
that do not dissipate.
High eel concentration would reduce the overlap parameter $\beta$, so the model does 
account for the sensitivity of the metachronal wave to eel concentration on the boundary.

Figure \ref{fig:omodel} shows that for $K<0.4$  (on the left side of the figure) entrained states are not present
at the end of the integration.   This is due
to groups of neighboring oscillators with initially 
large phase differences.   If integrated longer, these irregularities or discontinuities might eventually disappear.
The interaction parameter $K$ influences the time it takes for the short wavelength structure
to dissipate.  In a more realistic model, noise and diffusive interactions would also affect the range
of parameters giving an entrained or traveling wave state.
The odd black points at $\beta\approx 0.25, K =0.7$ are due to discontinuities at the left boundary
that continuously propagate through the system.  We are not sure why our left boundary condition
caused this problem only in this region of parameter space.

What properties of a phase oscillator model are required for a large basin of attraction to an entrained
or traveling wave state?
The model by \citet{Brumley_2012} is two dimensional as it depends
on oscillator radius as well as phase so it is more complex than a model that
consists only of a chain of phases.  With only phases, both our model and that by 
 \citet{Niedermayer_2008}
are not potential models, and interactions between pairs of oscillators are not applied equally and oppositely
to each oscillator in a pair, 
the way conventional physical forces are applied.    
  These three examples (our model,
and those by \citet{Niedermayer_2008,Brumley_2012})  of models 
developed
for traveling waves in biological systems might yield clues for more general classification
of the basins of attraction for phase oscillator models with local interactions.

For most of our integration parameters we saw only a single possible entrained state. 
Is it possible to predict the phase delay $\tau$, or wavelength, $N_\lambda$, of this entrained state? 
The integration 
shown in Figure \ref{fig:ov}a of the model given by Equation \ref{eqn:omodel}
shows that the phase at a single output time has
two regions,  One region has a low phase velocity and the other region has a higher 
phase velocity.     
In the fast and slow regions, the phase velocity is constant and phase differences between neighboring 
oscillators are maintained.  In appendix \ref{sec:entrained},  we  estimate the 
phase delay $\tau$ and wavelength $N_\lambda$ of the entrained state from the 
phase shifts that occur during the transitions between the fast and slow regions.
    
\begin{figure}
\includegraphics[width=3.4in, trim=15 0 0 0, clip]{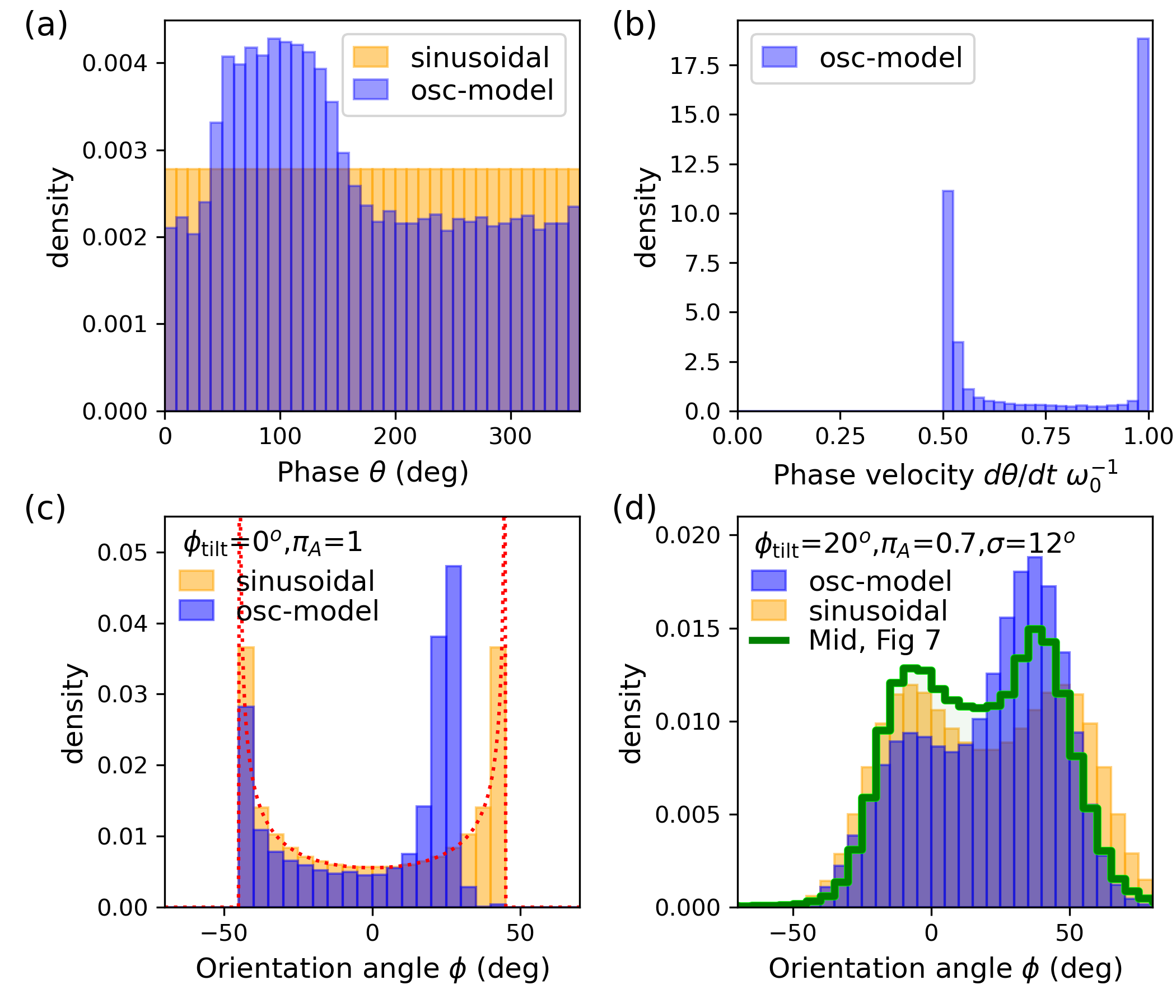}
\caption{Distributions for the directed chain integrated oscillator chain model shown in Figure \ref{fig:ov} are plotted in blue. 
These are compared to distributions for a constant phase velocity model which is shown in orange
and referred to as `sinusoidal'.
(a) The distribution of phase angles for the integrated oscillator chain model and the sinusoidal model.
(b) The distribution of phase velocities for the integrated oscillator chain model.  The sinusoidal model
has $d\theta/dt\ \omega_0^{-1} = 1$.
(c) The distribution of orientation angles for both oscillator chain and sinusoidal models computed using 
 equation \ref{eqn:phi}, $\phi_{\rm tilt}=0$ and $\pi_A  = A \omega_0/v = 1$. 
 The red dotted line shows the distribution function (in equation \ref{eqn:distp}) for
 the sinusoidal model.
(d) We show smoothed distributions of orientation angles 
 computed using equation \ref{eqn:phi}, $\phi_{\rm tilt}=20^\circ$ and $\pi_A =  0.7$
 for both oscillator chain and sinusoidal models. 
The sinusoidal and oscillator chain model distributions have been smoothed with a 
Gaussian filter with a standard deviation of $\sigma = 12^\circ$.
With a thick green line, we show the distribution of orientations measured
from the vinegar eels in \eelfour.  This
distribution is the same as plotted in green in Figure \ref{fig:hist}b.
The directed chain oscillator model displays an asymmetry in the associated 
orientation angle distributions (i.e., peaks of different heights) that is present in the 
observed distribution.
 \label{fig:omega_hist}
}
\end{figure}

\subsection{Distributions of orientation angles}
\label{sec:orient}

How do we relate the oscillator chain model to the orientation distributions displayed
in Figure \ref{fig:hist}b for the vinegar eels engaged in a metachronal wave?
The undulation velocity we measured in the freely swimming eel $v_u \sim 3.0$ mm/s is similar
to the metachronal wave velocity $v_{\rm MW} \sim 3.7 $mm/s so we could use
either one to make an estimate for how motions of the head propagate to the rest of the body. 
The free eel undulation frequency of $f_u = 5.9$ Hz gives intrinsic phase velocity 
$\omega_0 = 2 \pi f_u = 37\ {\rm s}^{-1} $.     
It is useful to compute the dimensionless ratio 
\begin{equation}
\pi_{A,{\rm MW}} \equiv 
\frac{A_{\rm MW} \omega_0}{v_{\rm MW}} \approx 0.70  \label{eqn:fac}
\end{equation}
using  parameters
listed in Table \ref{tab:free} and Table \ref{tab:meta} that we measured for the freely swimming eel
and metachronal wave.

The phase $\theta$ in our oscillator model represents the phase of back and forth oscillation  
in an eel's head.  We constructed our
interaction function assuming that the eel head moves away from
its mean centerline with coordinate perpendicular to the mean centerline $y = A \cos \theta $.
We assume that the head's motion excites a constant velocity traveling wave along
the eel body $y(x,t)$ with distance $y$ from the mean centerline a function 
of distance $x$ along the mean centerline.   The head's motion gives boundary condition 
\begin{equation}
y(x=0,t)  = A \cos \left[\theta(t)\right],  \label{eqn:bc}
\end{equation}
where the function $\theta(t)$ gives the phase of the head oscillation as a function of time.
With constant undulation wave velocity $v$
\begin{equation}
y(x,t)  = A \cos \left[\theta\left(t - \frac{ x}{v} \right)\right]  \label{eqn:yxt}
\end{equation}
is consistent with the boundary condition at $x=0$ (equation \ref{eqn:bc}).
The velocity that waves propagate down the eel body $v$ may not be the same as 
 $v_u$, the wave velocity for the freely swimming eel.

The slope of the body 
\begin{equation}
\frac{dy (x,t)}{dx} = A \sin \left[\theta \left(t -\frac{ x}{v} \right)\right]
\theta'  \left(t-\frac{ x}{v} \right) v^{-1} .
 \end{equation}
Here $\theta'$ is the derivative of the function $\theta(t)$.
The distribution of the slopes should be the same as the distribution of
$ \frac{A}{v} \frac{d \theta}{dt}  \sin \theta$ where the phases $\theta$ and phase velocities $\dot \theta$ 
are those at different times and positions for the heads in the oscillator array after the integration achieves
an entrained state. 
The slope of the body is $\frac{dy}{dx}= \tan\phi$ where $\phi$ is the body orientation angle.
From our model phases and phase velocities
we can compute the distribution of body orientation angles  $\phi$
assuming a constant wave velocity $v$ with 
\begin{equation}
\phi = \arctan \left[ \pi_A \left( \frac{d\theta}{dt} \frac{1} {\omega_0} \right)  \sin \theta \right]  + \phi_{\rm tilt},
\label{eqn:phi}
\end{equation}
with 
\begin{equation}
\pi_A \equiv  \frac{A \omega_0 }{v}.  \label{eqn:pi_A}
\end{equation}
We have purposely written equation \ref{eqn:phi} in terms of dimensionless parameters 
so as to facilitate comparison of our model with the vinegar eel collective motions.
Here the tilt angle $\phi_{\rm tilt}$, illustrated in Figure \ref{fig:overlap}, 
lets us adjust the angle of the eel centerlines with respect to the drop edge.

We generate model orientation distributions for the oscillator chain model with parameters and
integration shown in Figure \ref{fig:ov}.
In Figure \ref{fig:omega_hist} we use the arrays from 20 different times (spaced at 0.5 duration intervals)
to compute the distributions of phase angle $\theta$, phase velocity $\frac{d\theta}{dt}$,
and orientation angle $\phi$.  The orientation angles are computed
 with equation \ref{eqn:phi} from the phases and phase velocities.
The distributions have been normalized so that they integrate to 1. 
For comparison, we
 similarly generate and show distributions for a constant phase velocity model that
has $\frac{d\theta}{dt} = \omega_0$.  This model has a flat distribution of phases and can
be considered purely sinusoidal.
In this special case,
the orientation angle distribution function consistent with equation \ref{eqn:phi} and equation \ref{eqn:pi_A} is 
\begin{equation}
p(\phi)_{\rm sinusoidal} = \frac{1}{\pi} \frac{1 + \tan^2 (\phi - \phi_{\rm tilt})}{\sqrt{\pi_A^2 - \tan^2 (\phi - \phi_{\rm tilt} )}}. \label{eqn:distp}
\end{equation}

The phase velocity distribution for the oscillator chain model shown in Figure \ref{fig:omega_hist}b
shows two peaks, a low one for when there are interactions between neighboring oscillators 
and a high one that is at the intrinsic phase velocity.
This is what we would expect from inspection of the phase velocities in Figure \ref{fig:ov}.
Figure \ref{fig:omega_hist}c shows orientation angles $\phi$ computed with no tilt, $\phi_{\rm tilt} = 0$, 
and ratio $\pi_A =A \omega_0 /v =1$.    Orientation 
angle distributions for both oscillator chain model and constant phase velocity model exhibit two peaks and a trough. 
in Figure \ref{fig:omega_hist}c,  the constant phase velocity model distribution, in orange,  is consistent with the
distribution function of equation \ref{eqn:distp} that is shown with a dotted red line. 
The peaks of the orientation angle
distribution for the oscillator chain model have different heights due to the uneven phase velocity distribution, 
whereas the distribution is symmetrical about $\phi=0$ for the sinusoidal (constant phase velocity) model.

We can compare the modeled distribution
of body orientations to those measured in our videos of the eels engaged in the metachronal wave,  
 shown in Figure \ref{fig:hist}b,  and discussed in section \ref{sec:body_orient}.
Figure \ref{fig:omega_hist}d shows orientation angle distributions computed with $\phi_{\rm tilt} = 20^\circ$
and ratio $\pi_A = 0.7$ which is that of equation 
\ref{eqn:fac}.   To facilitate comparison between distributions
we have smoothed the model distributions using a Gaussian filter
with standard deviation of $12^\circ$.
In Figure \ref{fig:omega_hist}c, we replot one of the orientation angle distributions
that was shown in Figure \ref{fig:hist}b and is measured from  \eelfour \ of a metachronal wave.
The model orientation distributions shows two peaks, and when corrected
by the same factor (setting $\pi_A = \pi_{A,{\rm MW}}$) and smoothed, they have a width
and two peaks similar to that 
observed for the metachronal wave.    Unlike the observed distribution, the sinusoidal model's orientation
angle distribution 
is symmetrical about $\phi_{\rm tilt}$ and its two peaks are the same height.
In contrast, the oscillator chain model distribution is asymmetric or lopsided 
 and its two peaks have different heights. 
Because there are variations in oscillator phase velocity in the oscillator chain model,
the associated orientation angle distribution is lopsided.  
The oscillator chain model offers an 
explanation for the asymmetry that is present in the observed orientation angle distribution.

To compare the oscillator chain model to the observed orientation angle distribution we smoothed the model.
Noise-like variations in the observed orientation angle distribution can be due to eels that are not aligned with
their neighbors and variations in shading that affect the accuracy of the HOG algorithm.  
The oscillator chain model's distribution is more lopsided than the observed distribution which implies  
that variations in the phase velocity are not as extreme as predicted in Figure \ref{fig:omega_hist}b.
A more complex oscillator chain model would be needed to give  a better fit to the observed
orientation angle distribution.


\begin{figure*}
\includegraphics[width=6.5in, trim=0 0 0 0, clip]{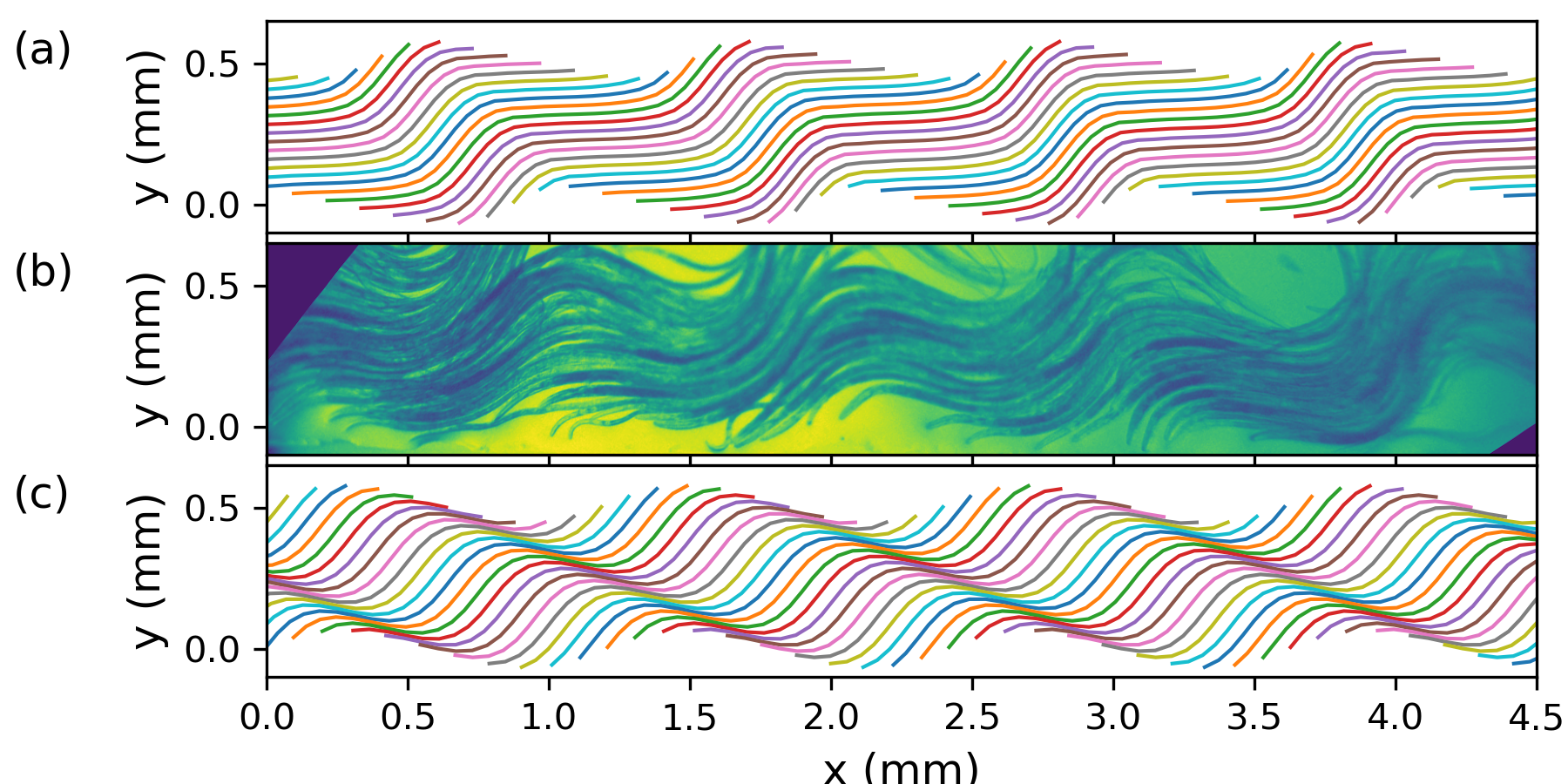}
\caption{(a)  Eel body positions that are computed with a series of outputs at different times 
of the phase oscillator
model shown in Figure \ref{fig:ov}  and using equation 
\ref{eqn:Xi}.  
Overlaps are reduced not only at the eel heads but throughout their body. 
(b) A panel from \eelfour \ similar to those shown in Figure \ref{fig:eels4}.
The morphology of the model wave in (a) resembles that seen in the vinegar eels.
(c) Eel body positions were estimated via equation \ref{eqn:Xi} but 
with a constant time delay and constant phase velocity model.
Other parameters were the same.
This model causes eel bodies to overlap.  A comparison between (a) and (c)
suggests that there must be variations in the phase velocities to reduce steric interactions.  
 \label{fig:wiggle}
}
\end{figure*}

\subsection{Body shapes}
\label{sec:shape}

In equation \ref{eqn:phi} we used model phases 
to compute the distribution of body orientation angles  $\phi$
assuming a constant wave velocity $v$.  With the same assumption
we can compute the position and shape of the entire body using a time series of model outputs. 
Our procedure for doing this is described in appendix \ref{sec:Xi}.

In Figure \ref{fig:wiggle}a we show computed eel body shapes that are derived
from the integrated phase oscillator
model output shown in Figure \ref{fig:ov} (integrating equation \ref{eqn:omodel})
and computed along the body lengths using equation \ref{eqn:Xi}.   To generate the body positions we used
amplitude $A = 0.07$  mm,  (based on that measured from eel head motions
for eels engaged in the metachronal wave), and
intrinsic phase velocity $\omega_0 = 2 \pi f_u$ with $f_u = 5.9 $ Hz
based on freely swimming eels.  We adopted tilt angle $\phi_{\rm tilt} = 20^\circ$
(the same as we used to generate orientation distributions in Figure \ref{fig:omega_hist}).
To match the metachronal wavelength we used 
a horizontal distance between eel mean centerlines of $D= 0.11$ mm,  (as defined
in Figure \ref{fig:overlap}). 
Lastly we use a wave speed $v = 4.1$ mm/s. 
The ratio  $\pi_A = A \omega_0/v = 0.63$ is similar to given in equation \ref{eqn:fac}
and was used to create the model orientation distributions in  Figure \ref{fig:omega_hist}d.
The eel body shapes using these parameters 
are shown in Figure \ref{fig:wiggle}a and they illustrate similar morphology to 
the vinegar eels themselves when engaged in the metachronal wave. 
Figure \ref{fig:wiggle}b shows a panel like those of Figure \ref{fig:eels4} from \eelfour \ for
comparison.

Figure \ref{fig:wiggle}a shows that 
the periodic variations in phase delay and phase velocity of an entrained state from 
 our oscillator chain model (equation \ref{eqn:omodel}) reduce overlap between eels, not just near
the eel heads but throughout their bodies.  The eel bodies are nearly equidistant from each other everywhere.
In Figure \ref{fig:wiggle}c we show
body positions generated with a constant phase velocity ($\omega_0$)
and constant phase delay (with the same wavelength $N_\lambda$) model.
The constant phase delay and phase velocity model fails badly.
Variations in phase delay between neighboring 
eels and in their phase velocity during different parts of the oscillation are probably needed 
to prevent strong steric interactions between the eels.


We chose the wave speed $v$ along the body to best match the observed morphology, however
it exceeds both the metachronal wave
speed of about $v_{\rm MC} \sim 3.7$ mm/s and the undulation wave speed on the 1 mm long freely swimming
eel of $v_u \sim 3.0 $ mm/s.   
 We might expect $v = v_{\rm MC}/\cos \phi_{\rm tilt} =
 3.9 $ mm/s using $v_{\rm MC} =3.7$ mm/s and $\phi_{\rm tilt} = 20^\circ$. 
 Our chosen value for $v$ exceeds this.  
Our assumption for computing orientation angle $\phi$ in equation \ref{eqn:phi} and body
shape ignores  
 interactions between organisms that should affect the speed of wave propagation down the eel bodies.
A more complex model that takes into account proprioception feedback throughout 
 the eels body lengths might give a smoother and more symmetric orientation angle distribution,
 (reducing the discrepancy between that modeled and measured  
 in Figure \ref{fig:omega_hist}d) 
 and a closer match to the wave morphology (improving the comparison
 between Figure \ref{fig:wiggle}a and b).  
 We observe that the amplitude of motion in the metachronal wave $A_{\rm MW} > A_u$
 exceeds the amplitude of undulation when freely swimming, $A_{\rm MW} > A_u$ and
 the speed of waves traveling down the body exceeds that when freely undulating $v > v_u$.
 A feedback motor control model, perhaps based on local body curvature, might predict or 
 explain these characteristics.
 
There is a discrepancy between the overlap parameter $\beta =d/A= 0.25$
 of the numerical oscillator model we adopted (shown in Figure \ref{fig:ov} and used to create 
 Figures \ref{fig:omega_hist} and \ref{fig:wiggle}) and that derived from the additional parameters we used to make Figure \ref{fig:wiggle}a for the eel bodies.
The distance between eel centerlines $d$ is related
 to the horizontal distance between mean centerlines $D$ 
 with $d = D \sin \phi_{\rm tilt}$ (see Figure \ref{fig:overlap}). 
For the model shown in Figure \ref{fig:wiggle}a, we used $D=0.11$ mm, $A = 0.07$ mm 
and $\phi_{\rm tilt} = 20^\circ$ giving $d = 0.038$ mm.  
We can estimate an overlap parameter for the tilted system 
 $\beta \sim D \sin \phi_{\rm tilt}/A = 0.54$ which exceeds our oscillator model overlap.
This discrepancy might be reduced if we included the eel body width 
and the tilt angle $\phi_{\rm tilt}$ in our overlap criterion function.
A more complex model that takes into account feedback throughout 
 the eels body lengths might also resolve this discrepancy.

\section{Summary and Discussion}
\label{sec:sum}

We presented high speed videos of swimming vinegar eel nematodes ({\it T. aceti})
 at low and high concentration.   
  In a drop containing a  high concentration of the vinegar eels, the eels concentrate 
at the edge of a drop and engage in collective wave-like motion known as a metachronal wave.
We found that freely swimming organisms have oscillation frequency
of about 6 Hz. 
However, at high concentration
the nematodes cluster on a boundary and exhibit traveling waves with a lower frequency of 
about 4 Hz. 
For a freely swimming vinegar eel, the body shape is nearly sinusoidal over much of its body length.
In contrast, the distribution of body orientation angles for organisms engaged in the metachronal
wave has two peaks of different heights, implying that the motion is
not purely sinusoidal.   The bodies spend more time
at higher  orientation angles w.r.t to their mean body orientation angle (averaged over a cycle).

We constructed a model for the collective behavior based on a chain of phase 
oscillators.    Because we do not see large drifts in the mean eel head positions, averaged over
an oscillation cycle, we neglect the head's forward motion.  Because
experiments of a similar nematode, {\it C. elegans}, support a model where the undulation is initiated 
at the head \citep{Wen_2012}, we use the phase of the head's back and forth motion to describe it as an oscillator.
Because the metachronal wave frequency is lower than the undulation frequency of a freely swimming eel,
we adopt interactions that reduce the oscillator phase velocity.  
Our oscillator model uses strong but directed or one-sided nearest neighbor 
to mimic steric interactions between organisms.  
The oscillator model (equation \ref{eqn:omodel}) robustly
exhibits entrained or traveling wave solutions and can have traveling waves with wavelength
(in terms of numbers of organisms or oscillators)
and mean phase velocity (in units of the intrinsic or freely swimming undulation frequency) similar to 
that of the vinegar eels when engaged in a metachronal wave.

To estimate the distribution of body orientation angles and body shapes from our oscillator model,  
we assume that the undulation waves propagating down the body from the eel head 
have a constant wave velocity.   This gives a 
 two humped distribution of body orientations with peaks of different heights, similar to that observed for vinegar eels 
 engaged in the metachronal wave.   The body shapes are similar 
 to those engaged in the wave and the eel bodies don't overlap
 over their entire length.  The model which was designed to impede eel head overlaps
 also reduces close interactions throughout the eel bodies.   
 
Our model neglects interactions between organisms that should affect the amplitude and speed of 
wave propagation down the eel bodies.   
Our model also neglects the ability of the eels to change direction and congregate.   
Improved models could take into account the positions and phases of all points in the eel's bodies and allow
them to swim, reorient and congregate. 
 
Few known simple phase oscillator models exhibit a large basin
of attraction to an entrained or traveling wave state. 
Perhaps our model (given in equation \ref{eqn:omodel}) and that by 
\citet{Niedermayer_2008} can serve as examples that might give insight for 
more general classification of coupled phase oscillator models that would be helpful for predicting wavelike
collective behavior.

Vinegar eels are visible by eye and are large compared to other biological systems
that exhibit metachronal waves, such as carpets of cilia \citep{Tamm_1970,Tamm_1975,Elgeti_2013}
or flagella on the surface of {\it Volvox carteri} alga colonies \citep{Brumley_2012}.  Their large size
facilitates study, however it also places them in an interesting intermediate  
hydrodynamic regime, with swimming Reynolds number $Re  = v_{\rm swim} L/\nu
\sim 0.4$  (where $\nu \sim 1\ {\rm mm}^2 {\rm s}^{-1}$ is the kinematic viscosity of water), 
so the nature of hydrodynamic interactions between 
them should differ from that of microorganisms which are at much lower Reynolds number
(e.g., \citep{Elgeti_2015,Gilpin_2020}).   
Their proximity when involved in collective behavior suggests
that steric interactions may be important.  Studies of the similar nematode {\it C. elegans} locomotion \citep{Wen_2012}  imply that feedback in motor control affects their gait.    It is exciting to have a 
relatively large system in which collective motion can be studied, however, this system also presents
new challenges for understanding its behavior.

In on-going studies we will describe experiments of concentrations
of {\it C. elegans}, explore collectively formed dense coherent filaments in {\it T. aceti} that we have
observed advance on a vinegar/oil interface and
explore the role of concentration, drop shape and wetting angle in affecting metachronal wave formation
in {\it T. aceti} \citep{Peshkov_2021}.
Similarities between {\it T. aceti} and {\it C. elegans} suggest that it may be possible to use techniques 
developed for {\it C. elegans} to perform genetic modifications on the {\it T. aceti} nematode.  
In future, genetically modified strains
may help us better understand the molecular underpinnings of the collective motion.  
Future studies could question whether there is an evolutionary advantage to  the 
collective behavior which may help populations of nematodes penetrate crowded environments 
to reach food or drive flows that transport oxygen and nutrients.  

\begin{acknowledgments}

We thank Ed Freedman for the gift of a microscope. We thank Kanika Vats
for advice with filming through a microscope and letting us try filming at high speed 
with an inverted microscope.  
We thank William Houlihan for lending us an inverted microscope. 
We thank Nick Reilly for obtaining a centrifuge and showing us how to us it.
We thank Doug Portman for helpful discussions on {\it C. elegans}.
We thank Keith Nehrke, Sanjib K. Guha, Yunki Im and other members of 
Nehrke's lab for helping us explore  {\it C. elegans}, teaching
us how to culture {\it C. elegans} and giving us some materials and live worms to culture in our lab. 
We thank Steve Teitel, Sanjib K. Guha, Keith Nehrke and Randal C. Nelson for helpful 
suggestions and discussions.

This material is based upon work supported in part  by NASA grants 80NSSC21K0143 and 80NSSC17K0771, 
National Science Foundation Grant No. PHY-1757062, and  
National Science Foundation Grant No. DMR-1809318.

\end{acknowledgments}


\bibliography{eels}

\appendix

\begin{figure}
\centering
\includegraphics[width=3.3in, trim=10 5 0 -10, clip]{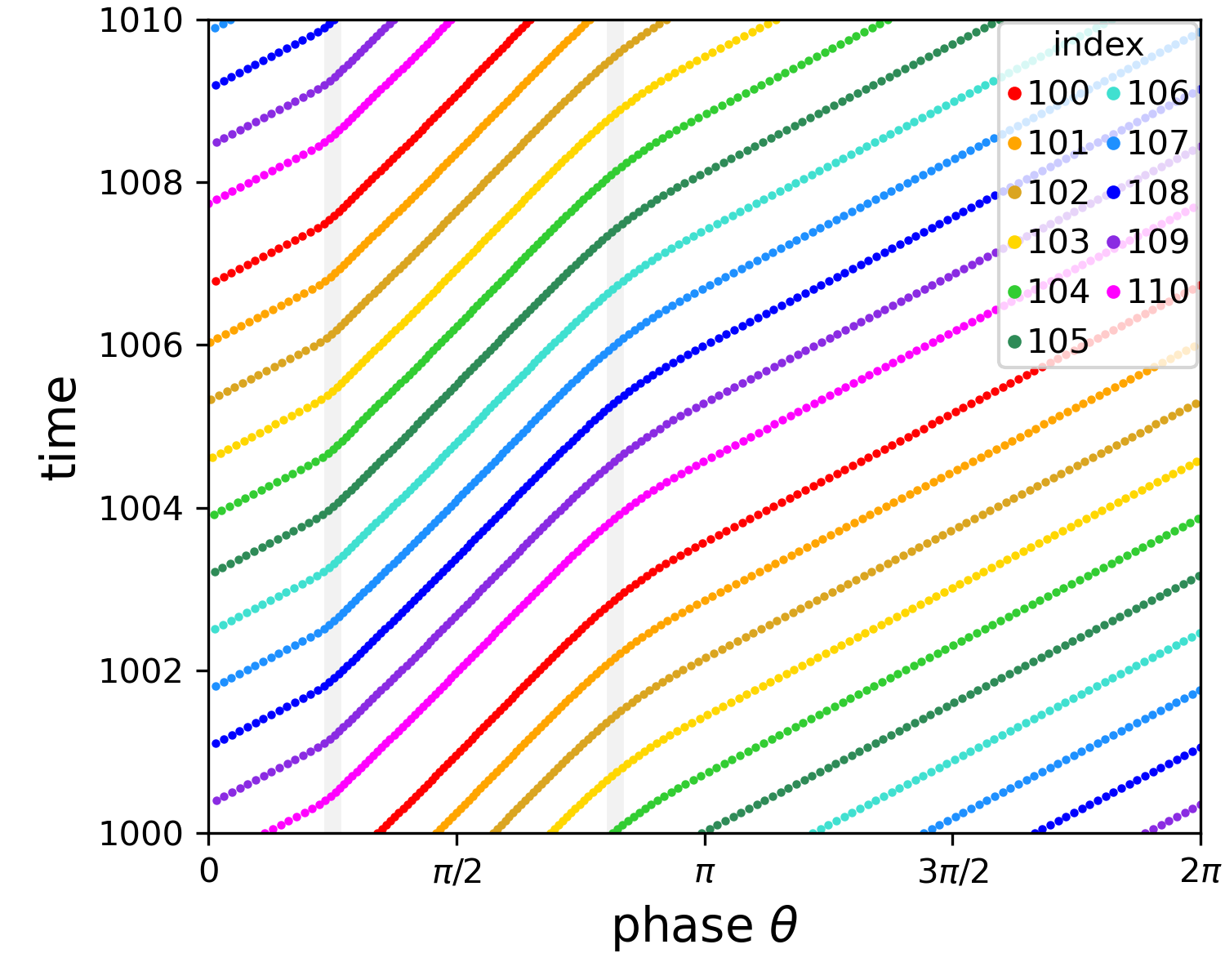}
\caption{  
We plot phase $\theta$ vs time for 11 consecutive oscillators for the directed chain oscillator model with the same
parameters as shown in Figure \ref{fig:ov}. 
Each oscillator is plotted with a different color and the oscillator indices are given in the key.  
The figure shows the periodic compression and rarefaction of phase in the entrained state.  The region
of lower phase velocity, between $\theta \approx  0.25\pi$  and  $0.82 \pi$,  is marked with
the vertical thick light gray lines. The inverse of the local slope of one of the curves 
gives the phase velocity and the horizontal distance between neighboring curves gives the phase delay
between consecutive oscillators.
\label{fig:char}}
\end{figure}

\section{Compression and rarefaction in entrained states}
\label{sec:entrained}

The integration 
shown in Figure \ref{fig:ov} of the model given by Equation \ref{eqn:omodel}
shows that each oscillator has
a periodic trajectory but with two regions.  One region has a low phase delay 
and phase velocity and the other region has a higher phase delay and
phase velocity.     
We can also show this behavior 
by plotting phase angle $\theta$ against time for a series of oscillators.  This type of plot
is often used to study shock compression or rarefaction. On this plot, the inverse of the  slope 
gives the phase velocity and the horizontal distance between consecutive lines gives the phase delay.
We show such a plot in Figure \ref{fig:char} for an integration with the same
parameters as in Figure \ref{fig:ov}. We plot phase $\theta$ vs time for 11 consecutive oscillators
after integrating to $t=1000$ and for a duration of $\Delta t=10$.  
The region of lower phase velocity lies between the thick gray vertical lines which are at
$\theta=0.25 \pi$ and $0.82 \pi$.

We make the assumption that an entrained state has two regions, like those seen at the end of the integrations
shown in Figure \ref{fig:ov} and in Figure \ref{fig:char}.  For our model in equation \ref{eqn:omodel},
which we repeat here for clarity,
\begin{equation}
\frac{d \theta_i}{dt} \omega_0^{-1} =  1 - \frac{K}{2} 
\left[ \tanh \left(  \frac{\cos \theta_{i-1} - \cos \theta_i -  \beta }{ h_{ol} }\right) + 1\right] . 
\label{eqn:omodel2}
\end{equation}
the high value of the phase velocity is the intrinsic phase velocity $\omega_0$
and the low value is $\omega_0 (1 - K)$.
An entrained state has a phase delay $\tau$ where 
\begin{equation} 
\theta_j (t + \tau) = \theta_{j+1} (t) . \end{equation} 
We expand the left side to first order in $\tau$ and write $\theta_{j+1}$ in terms
of the phase delay $\chi_j = \theta_{j+1} - \theta_j$, giving
\begin{equation}
\dot \theta_j (t) \tau \approx \chi_j . \label{eqn:ttau}
\end{equation}
We denote the phase delay for the slower state as $\chi_s$ and that 
of the faster state as $\chi_f$.
Equation \ref{eqn:ttau} gives 
\begin{align}
\chi_s &\approx \omega_0(1-K) \tau \nonumber \\
\chi_f &\approx \omega_0  \tau. \label{eqn:chisf}
\end{align}

In the fast and slow regions, the phase velocity is constant and phase differences between
oscillators are maintained.    The properties of the entrained states must be
set by the transition regions.
We consider two oscillators, one in the slow region and the other that is exiting
the slow region.  
We can estimate the change in phase delay between the two regions from the time $\Delta t_{fs}$ it takes 
 a single oscillator to exit the slow region
 \begin{equation}
 \chi_f - \chi_s \approx \Delta t_{fs}\  \tilde \omega,  \label{eqn:change}
 \end{equation}
 where 
 \begin{equation}
 \tilde \omega \approx (1-K/2) \omega_0  \label{eqn:tildeo}
 \end{equation}
 is the average phase velocity.
We use equation \ref{eqn:change} to estimate the phase delay $\tau$.

For small phase delay $\chi_{j-1} = \theta_j - \theta_{j-1}$
equation \ref{eqn:omodel2} can be written to first order in phase shift $\chi_{j-1}$ as 
\begin{equation}
\frac{ d\theta_j }{dt} \omega_0^{-1}  \approx 1  - \frac{K}{2}
 \left[  \tanh \left( \frac{\sin \theta_j\  \chi_{j-1} -\beta}{h_{ol}} \right) + 1\right].
\end{equation}
The time $\Delta t_{fs}$ it takes oscillator $j$ to pass through the transition from
slow to fast regions we estimate
from the time it takes $|\sin \theta_j \chi_{j-1}|/h_{ol}$ to change by about 2  (corresponding
the region of high slope for the tanh function).  
This transition time  is approximately 
\begin{equation}
\Delta t_{fs} \sim 2h_{ol} \left|   \cos \theta_j \frac{d\theta_j}{dt} \chi_{j-1}  \right|^{-1}.
\end{equation}
We assume that the transition boundaries are where $|\cos \theta| \sim 1$  
and take an average of the fast and slow values for $\frac{d\theta}{dt}$ and $\chi$ 
(using equations \ref{eqn:chisf} and \ref{eqn:tildeo})
to estimate  the duration of the transition from a fast to slow region or vice versa
\begin{align}
\Delta t _{fs} &\sim \frac{8 h_{ol}} {(\chi_f + \chi_s)( 2- K) \omega_0}  .
\end{align}
Using equations \ref{eqn:change} and  \ref{eqn:chisf}
we estimate the delay $\tau$
\begin{equation}
\tau \sim \frac{2}{\omega_0} \sqrt{\frac{h_{ol}}{K(2 - K)}}
\end{equation}
and the wavelength 
\begin{equation}
N_\lambda \sim \frac{ 2 \pi}{\tilde \omega \tau} \sim 2 \pi \sqrt{\frac{K}{(2-K)h_{ol}}} . \label{eqn:Nl}
\end{equation}
For $h_{ol} = 0.05$ and $K=0.5$ this gives $N_\lambda \sim 16$ 
 which is a reasonable value but exceeds the value
of 12 we see in the integration shown in Figure \ref{fig:char}.  
We verified that $N_\lambda$ decreases with increasing $h_{ol}$, though it does not decrease
as quickly as predicted by equation \ref{eqn:Nl}.   A better prediction would take into account
the phases of the transitions and the difference between compression and rarefaction transitions.
The comparison between estimated and numerically measured wavelengths suggests that techniques 
 used to study non-linear differential 
 equations may be useful for predicting the properties of entrained states.
 
\section{Predicting body positions and shapes from a phase oscillator model}
\label{sec:Xi}

In this section we show how we compute eel body shapes and positions from 
an oscillator chair model.  We assume the eel head positions
are described by a chain of oscillators as illustrated in 
Figure \ref{fig:overlap}.   We assume that waves propagate down
the body with a constant speed $v$.  
 
We adopt an Cartesian coordinate system ${\bf X} = (X,Y)$ on the plane
to describe positions of points on the body of
a chain of eels, as shown in Figure \ref{fig:overlap}b. 
We assume the mean centerline position of the $i$-th eel's head has coordinates ${\bf X}_{i,hc}$
and the eel's mean centerline is tilted by angle $\phi_{\rm tilt}$ with respect to the horizontal direction.
We assume that the mean centerline head positions are fixed and are equally spaced on the $X$ axis
\begin{equation}
 {\bf  X}_{i,hc}  = \begin{pmatrix} 
 i D  \\ 0   \end{pmatrix} , \label{eqn:Xhead}
\end{equation}
where $D$ is the horizontal distance between the mean centerlines.
We assume the wave travels down the body with velocity $v$,  
as given in equation \ref{eqn:yxt} which we repeat here 
\begin{equation}
y_i(x,t) = A \cos \left[ \theta_i \left( t - \frac{x}{v} \right)  \right].  \label{eqn:yagain}
\end{equation}
The $i$-th eel's head position is at $y(x=0,t)$.  Here $x$ is the distance along the mean centerline
and $y_i$ is the distance perpendicular to it.
We rotate the centerlines by $\phi_{\rm tilt}$ so that in the $(X,Y)$ coordinate system 
the head of the $i$-th eel is at 
\begin{align}
{\bf X}_{i,h} (t)  &= 
\begin{pmatrix} 
		\cos \phi_{\rm tilt} &- \sin \phi_{\rm tilt} \\
		\sin \phi_{\rm tilt} & \cos \phi_{\rm tilt} 
	\end{pmatrix} 
 \begin{pmatrix} 
	0  \\ A \cos (\theta_i(t)) 
\end{pmatrix}  \nonumber \\
& \ \ \ \ + {\bf X}_{i,hc} 
.\qquad \qquad \qquad  \label{eqn:Xih}
\end{align}
We can use the coordinate along the mean centerline $x$ to specify body positions 
\begin{align}
 {\bf X}_i(x,t) 
&=  \begin{pmatrix}  
		\cos \phi_{\rm tilt} &- \sin \phi_{\rm tilt} \\
		\sin \phi_{\rm tilt} & \cos \phi_{\rm tilt} 
	\end{pmatrix}
 \begin{pmatrix} 
	x \\ A \cos \left[\theta_i\left(t - \frac{x}{v}\right)\right] 
 \end{pmatrix} \nonumber \\
&  \ \ \ \ +    {\bf X}_{i,hc}  . \qquad \qquad \qquad \label{eqn:XX}
\end{align}
With $x=0$, this is consistent with equation \ref{eqn:Xih} for the $i$-th eel's head.

Using equation \ref{eqn:yagain}, 
at $t=0$, the $y$ position of the $i$-th eel is determined by its head position at an earlier time,
\begin{align}
 y_i(x,t=0) &= A \cos \left[ \theta_i \left( -\frac{x}{v} \right) \right] ,
 \end{align}
 where the earlier time is 
 \begin{equation}
 t' = -\frac{x}{v} . \label{eqn:tprime}
 \end{equation}
Using a phase oscillator model we can generate arrays of phases $\theta_i$ at a series of times.
The arrays at different output times $t'$ then can be used 
 to predict the ${\bf X} $ positions at $t=0$ along the eel's bodies; 
\begin{align}
{\bf X}_i(t')  
&=  \begin{pmatrix} 
		\cos \phi_{\rm tilt} &- \sin \phi_{\rm tilt} \\
		\sin \phi_{\rm tilt} & \cos \phi_{\rm tilt} 
	  \end{pmatrix} 
	   \begin{pmatrix} 
	-vt' \\ A \cos \left[\theta_i\left( t'\right)\right] 
	\end{pmatrix} 
 \nonumber \\
&  \ \ \ \ +  \begin{pmatrix}   iD \\ 0     \end{pmatrix}, \qquad \qquad \qquad  \label{eqn:Xi}
\end{align}
where we have used equations \ref{eqn:Xhead}, \ref{eqn:XX}  and \ref{eqn:tprime}.

From a series of phase arrays computed at different output times for the phase oscillator model
of equation \ref{eqn:omodel} we can generate eel body positions using equation \ref{eqn:Xi}.
To do this we require values for the velocity of waves along the eel body $v$,
the amplitude $A$,  the horizontal distance between the mean positions of organism heads $D$
and the body tilt angle $\phi_{\rm tilt}$.  Also, the outputs of the integration must be put in units
of time using the intrinsic oscillator phase velocity $\omega_0$.

\end{document}